\documentclass[prl,twocolumn,aps,showpacs,amsmath,amssymb,superscriptaddress]{revtex4-1}

\usepackage{color}
\usepackage{bm}
\usepackage{graphicx}
\usepackage{braket}

\usepackage[latin9]{inputenc}
\usepackage{color}
\usepackage{amsmath}
\usepackage{amssymb}
\usepackage{esint}
\usepackage[unicode=true,
 bookmarks=true,bookmarksnumbered=false,bookmarksopen=false,
 breaklinks=false,pdfborder={0 0 1},backref=false,colorlinks=true]
 {hyperref}
\hypersetup{
 linkcolor=magenta, urlcolor=blue, citecolor=blue, pdfstartview={FitH}, hyperfootnotes=false, unicode=true}
\usepackage{color}

\newcommand{\bea}{\begin{eqnarray}}
\newcommand{\eea}{\end{eqnarray}}

\newcommand{\be}{\begin{equation}}
\newcommand{\ee}{\end{equation}}



\begin{document}
\title{Transport properties across the many-body localization transition in quasiperiodic and random systems} 
\author{F. Setiawan}
\author{Dong-Ling Deng}
\affiliation{Condensed Matter Theory Center and Joint Quantum Institute, Department of Physics, University of Maryland, College Park, MD 20742, USA}
\author{J. H. Pixley}
\affiliation{Condensed Matter Theory Center and Joint Quantum Institute, Department of Physics, University of Maryland, College Park, MD 20742, USA}
\affiliation{Department of Physics and Astronomy, Center for Materials Theory, Rutgers University, Piscataway, NJ 08854, USA}

\begin{abstract}
We theoretically study transport properties in one-dimensional interacting quasiperiodic systems at infinite temperature. We compare and contrast the dynamical transport properties across the many-body localization (MBL) transition in quasiperiodic and random models. Using exact diagonalization we compute the optical conductivity $\sigma(\omega)$ and the return probability $R(\tau)$ and study their average low-frequency and long-time power-law behavior, respectively. 
We show that the low-energy transport dynamics is markedly distinct in both the thermal and MBL phases in quasiperiodic and random models and
 find that the diffusive and MBL regimes of the quasiperiodic model are  more robust than those in the random system.
Using the distribution of the DC conductivity, we quantify the contribution of sample-to-sample and state-to-state fluctuations of $\sigma(\omega)$ across the MBL transition.
We find that the activated dynamical scaling ansatz works poorly in the quasiperiodic model but holds in the random model with an estimated activation exponent $\psi\approx 0.9$.
We argue that near the MBL transition in quasiperiodic systems,  critical eigenstates give rise to a subdiffusive crossover regime on finite-size systems.
\end{abstract}
\maketitle

\section{Introduction} 
Despite the presence of zero-point energy, it is possible to localize a quantum mechanical particle~\cite{Anderson-1958}. In single-particle problems, Anderson localization can occur due to either strong randomness~\cite{Abrahams-1979, Wegner-1980} or aperiodicity~\cite{Azbel-1979, Aubry-1980}. While both effects create exponentially localized single-particle wave functions and lead to Anderson insulating phases, they do so in a rather different manner. The insulating phase in the disordered problem is compressible with no gap in the single-particle spectrum and the optical conductivity is a smooth function of frequency which vanishes in the DC limit.
In contrast, a quasiperiodic  localized phase results from the multifractal gap structure of the single-particle Hamiltonian, which produces an optical conductivity that is not necessarily a smooth function of frequency. 

Recently, the development of many-body localization (MBL)  has generalized Anderson localization in both random~\cite{Basko_2006_AP, Oganesyan_2007_PRB,Imbrie_2016_JSP} and quasiperiodic~\cite{Iyer_2013_PRB, Schreiber-2015-Science,Mastropietro_2015_PRL} models to include interactions (for recent reviews see~\cite{2015_NH_review, 2015_Altman_review,Deng-2017-Many,Abanin-2017-recent}). The MBL phase transition
~\cite{Pal-2010, 2015_Potter_PRX, 2015_Vosk_PRX, Luitz_2015_PRB, Zhang-2016, Parameswaran-2016}
separates an ergodic (i.e., thermalizing) phase from a many-body localized phase (that never reaches thermal equilibrium). As a result, the MBL transition is inherently dynamical and is not a thermodynamic quantum phase transition, but is a transition in the many-body eigenstates. The thermal phase~\cite{Luitz-2016} is characterized by eigenstates that have a 
nonzero level repulsion, volume-law scaling of entanglement entropy, and satisfy the eigenstate thermalization hypothesis (ETH)~\cite{Deutsch_1991_PRA, Srednicki_1994_PRL, 2008_Rigol_Nature}. 
It is important to stress that the thermal phase is not necessarily a metal, as it does not have to support any DC transport~\cite{Lev-2015,2015_Agarwal_MBL_PRL,2015_Potter_PRX, 2015_Vosk_PRX,Khait-2016-Spin,Luitz-2016B,Agarwal-2017}.
On the other hand, the many-body localized phase has eigenstates that have no level repulsion (Poisson level statistics)~\cite{Oganesyan_2007_PRB}, area-law scaling of entanglement entropy~\cite{2013_Bauer_JSM} that grows logarithmically slow following a global quench~\cite{Bardarson-2012,Serbyn-2013a}, statistical orthogonality catastrophe \cite{Khemani-2015-Nonlocal, Deng-2015-Exponential}, and violates the ETH. In addition, the MBL phase is expected to have an ``emergent integrability,'' with an extensive number of local integrals of motion~\cite{Serbyn-2013b, Huse-2014,Imbrie-2017-Many}.

Similar to disordered quantum phase transitions~\cite{Vojta-2013}, the behavior near the MBL transition in random models is dominated by Griffith-McCoy effects,  where statistically rare configurations of the disorder potential create local regions in the system that are ``in the wrong phase''. These are nonperturbative \emph{sample-to-sample} fluctuations and can have dramatic effects in both the thermal and MBL phases~\cite{Agarwal-2017}. For example, Griffith effects are amplified in one dimension; on approach to the MBL transition from the thermal phase, the dynamics are expected to be dominated by local MBL regions of the system which act as insulating blocks that create bottlenecks for transport~\cite{2015_Agarwal_MBL_PRL,Agarwal-2017}. This leads to subdiffusion where the dynamical conductivity obeys a power law at low frequency $\sigma(\omega) \sim \omega^{\alpha}$ with $0 \le \alpha < 1$
 and $\sigma_{\mathrm{DC}} \sim 1/L^{z\alpha}$, 
 where $z$ is the dynamic exponent relating energy and length ($E \sim 1/L^z$). Based on scaling relations, one can show that these two exponents $z$ and $\alpha$ are related by $\alpha=1-2/z$ in the thermal phase~\cite{2015_Agarwal_MBL_PRL}. 
 As a result,  in the thermodynamic limit $\sigma_{\mathrm{DC}}$ vanishes in the Griffith regime within the thermal phase.  
 This picture is consistent with various MBL studies, such as the spreading of many-body wave packets using time-dependent density matrix renormalization group (tDMRG)~\cite{Lev-2015}, the optical conductivity within exact diagonalization (ED)~\cite{2015_Agarwal_MBL_PRL}, the nonequilibrium steady-state current calculations within tDMRG~\cite{Scardicchio-2016}, and strong-disorder renormalization group (RG) calculations~\cite{2015_Potter_PRX, 2015_Vosk_PRX,Parameswaran-2016}.
  The MBL transition is accompanied by infinitely slow relaxation, where $z\rightarrow \infty$ (Refs.~\cite{2015_Potter_PRX, 2015_Vosk_PRX}), which gives rise to $\sigma(\omega) \sim \omega$ (Ref.~\cite{2015_Agarwal_MBL_PRL}) and broad distributions of  observables with long tails~\cite{Luitz-2016C, Yu-2016}. Deep in the MBL phase, the system is an insulator with a vanishing $\sigma_{\mathrm{DC}}$ ~\cite{Berkelbach-2010,Prelov-2010} and $\alpha \rightarrow 2$~\cite{2015_Gopalakrishnan_PRB}.
On the other hand, nonperturbative effects in the MBL phase arise due to local thermal rare regions~\cite{2015_Gopalakrishnan_PRB,Agarwal-2017} that can entangle with and thermalize the neighboring regions around themselves. It has been argued that these rare thermal regions can grow and thermalize the entire system in dimensions greater than one~\cite{DeRoeck-2016}. 

The physics of MBL in quasiperiodic systems remains largely unexplored compared with its random counterpart, but is starting to gain considerable (and necessary) attention. One interesting, yet seemingly obvious consequence of quasiperiodic potentials is the absence of randomness, which implies that there should be no Griffith effects. This has numerous consequences, one of which suggests that MBL in quasiperiodic models is more robust than that in random models, a question that we will explore in this paper. 
 The random phase of the incommensurate potential can be thought of as a correlated disorder potential that is the same at each site for each sample. 
In addition to these long-range correlated sample-to-sample deviations,
there are also fluctuations over eigenstates that are all weighted equally at infinite temperature.
While quasiperiodic MBL is of fundamental interest in its own right, quasiperiodic potentials offer the chance to study the effect of single-particle mobility edges on MBL \cite{Li-2015-Many,Modak-2015-Many,Li-2016-Quantum,Li-2017mobility}, can host a form of localization protected order \cite{Huse-2013-Localization,Chandran2017localization}, and can be realized in ultracold atom experiments \cite{Schreiber-2015-Science}.  Unfortunately, there are a lack of analytic tools available to study the transition for quasiperiodic potentials since many methods rely on a random distribution for the couplings such as real space strong disorder RG. 
Recent numerical work~\cite{Khemani-2017} has shown that near the MBL transition, the variance of the entanglement entropy in the quasiperiodic model is dominated by fluctuations of eigenstates, whereas the random MBL transition is dominated by sample-to-sample fluctuations. 
How the dichotomy between sample versus eigenstate fluctuations dictates the nature of transport near the MBL transition in quasiperiodic models is not well understood. 

Due to these considerations, it is somewhat surprising that the numerical data of the level statistics and entanglement entropy across the MBL transition in quasiperiodic and random models look so similar. For example, exact diagonalization studies in the random model~\cite{Luitz_2015_PRB} and in the quasiperiodic model~\cite{Khemani-2017} have both found a correlation length exponent $\nu \approx 1$. Despite this similarity, due to general distinctions between randomness and quasiperiodicity these two results have different implications. For the random model, this result ($\nu \approx 1$) 
violates the Chayes-Chayes-Fisher-Spencer (CCFS) criteria \cite{Chayes-1986-Finite} and a many-body generalization of the Harris criteria~\cite{Chandran-2015-Finite} ($\nu\geq 2/d$). On the other hand, the results for the quasiperiodic model~\cite{Khemani-2017} are consistent with a Harris-Luck  criteria \cite{Harris-1974-Effect,Luck-1993-Classification} ($\nu \ge 1/d$). It is not obvious what quantities will clearly distinguish MBL in quasiperiodic and random models. One natural place to look is the transport properties since the Griffith's picture provides a description of the existing numerical data in random models. It is an interesting question to ask whether the transport data is \emph{qualitatively} different in the presence of quasiperiodic potentials. Central to this question is the behavior of the dynamic exponent $z$ across the quasiperiodic MBL transition. In the MBL phase, $z=\infty$ as the system is an insulator independent of the nature of the potential. Therefore, going across the thermal-to-MBL transition without any Griffith effects, it is not clear if $z$ will diverge like a power law similar to the random model, jump discontinuously to $\infty$ across the transition, or do something else entirely. The physical mechanisms dictating the low-energy transport properties near the MBL transition in quasiperiodic models is an interesting and open question that we address in this paper.

In this paper, we study the transport properties in the one-dimensional (1D) interacting Aubry-Andre (AA) model \cite{Aubry-1980} at infinite temperature across the MBL transition. Using exact diagonalization, we compute the optical conductivity using Kubo formula, its DC limit, and the return probability. We compare and contrast these calculations for the AA model with those on a random generalization of the AA model (with a random phase at each \emph{site}). Our results show that the thermal and MBL regimes  are
markedly distinct between the random and quasiperiodic models where both the diffusive thermal regime and MBL phase are more robust in the quasiperiodic model, e.g.,
the MBL phase in the quasiperiodic model is a much better insulator compared with its random counterpart.
We systematically compare the sample-to-sample and state-to-state fluctuations in the transport properties of the random and quasiperiodic models.
Our data for the quasiperiodic model are consistent with a subdiffusive crossover regime that shrinks with increasing system size and does not seem to obey activated dynamical scaling. In contrast, our data for the random model displays activated dynamical scaling with an estimated activation exponent $\psi \approx 0.9$ (in good agreement with the RG predictions~\cite{2015_Potter_PRX, 2015_Vosk_PRX} of $\psi=1$) on approach to the thermal-to-MBL transition. We use our numerical data to construct a schematic crossover diagram for the transport properties near the MBL transition in quasiperiodic systems and argue that the quantum critical crossover regime gives rise to subdiffusion in finite-size systems. 

This paper is organized as follows. We begin by introducing the AA and random models in Sec.~\ref{sec:model}. We then show the disorder-averaged level statistics and half-chain entanglement entropy used in estimating the location of the MBL transition. In Sec.~\ref{sec:transport}, we study the transport properties across the phase diagram by comparing and contrasting the DC conductivity, optical conductivity, and return probability for the AA and random models.  In Sec.~\ref{sec:sigmadcdist}, we study the distributions of the DC conductivity across the phase diagram and present a detailed comparison between the quasiperiodic and random models. In this section, we also quantify the contribution of sample-to-sample and state-to-state fluctuations to the transport. In Sec.~\ref{sec:scenario}, we use the activated dynamical scaling ansatz to compare the optical conductivity near the MBL transition of the AA and random models and also present a scenario for the nature of transport in the quasiperiodic thermal-to-MBL transition.  We end with a conclusion in Sec.~\ref{sec:conclusion}.

\section{Models}\label{sec:model}
We study the 1D interacting AA and random models, which are both defined as
\begin{equation}
H = \sum_j t (c_{j+1}^{\dag}c_j + \mathrm{H.c.}) + V_j n_j + U n_jn_{j+1},
\label{eqn:ham}
\end{equation}
where the density operator $n_j = c_j^{\dag} c_j$, $t$ is the hopping strength, $U$ is the nearest-neighbor interaction, and the potential term $V_j$ can be either quasiperiodic or random. In this paper, we set $t = 1$ and $U = 0.5t$. For the quasiperiodic model, we take $V_j^{\mathrm{AA}} =  \lambda \cos(2 \pi b j + \phi)$ with a random phase $\phi \in [0,2\pi)$ that is the same at each site and $b=2/(1+\sqrt{5})$ which is an irrational number. We note that our results do not depend qualitatively on the exact value of $b$ as long as $b$ is an irrational number. For the random model, following Ref.~\cite{Khemani-2017} we consider $V_j^\mathrm{R} =  \lambda \cos(2 \pi b j + \phi_j)$ with a random phase at each site $\phi_j \in [0,2\pi)$. This is a natural random generalization of the AA potential where each site has the same distribution of the potential with the distinction from the quasiperiodic model being that the sites here are not correlated. One major advantage of using this distribution is that it allows us to compare data from the two models at the same $\lambda$.
We take periodic boundary conditions (unless otherwise stated) for system size $L$ and focus on half filling. We use ED to compute the eigenstates, and focus on the states in the middle third of the many-body spectrum. The number of random samples used ranges from 60,000 ($L$ = 10) to 2,500 ($L$ = 16). In the following, we refer to the quasiperiodic model as AA and the random model as random.

\begin{figure}[h!]
\begin{center}
\includegraphics[width=\linewidth]{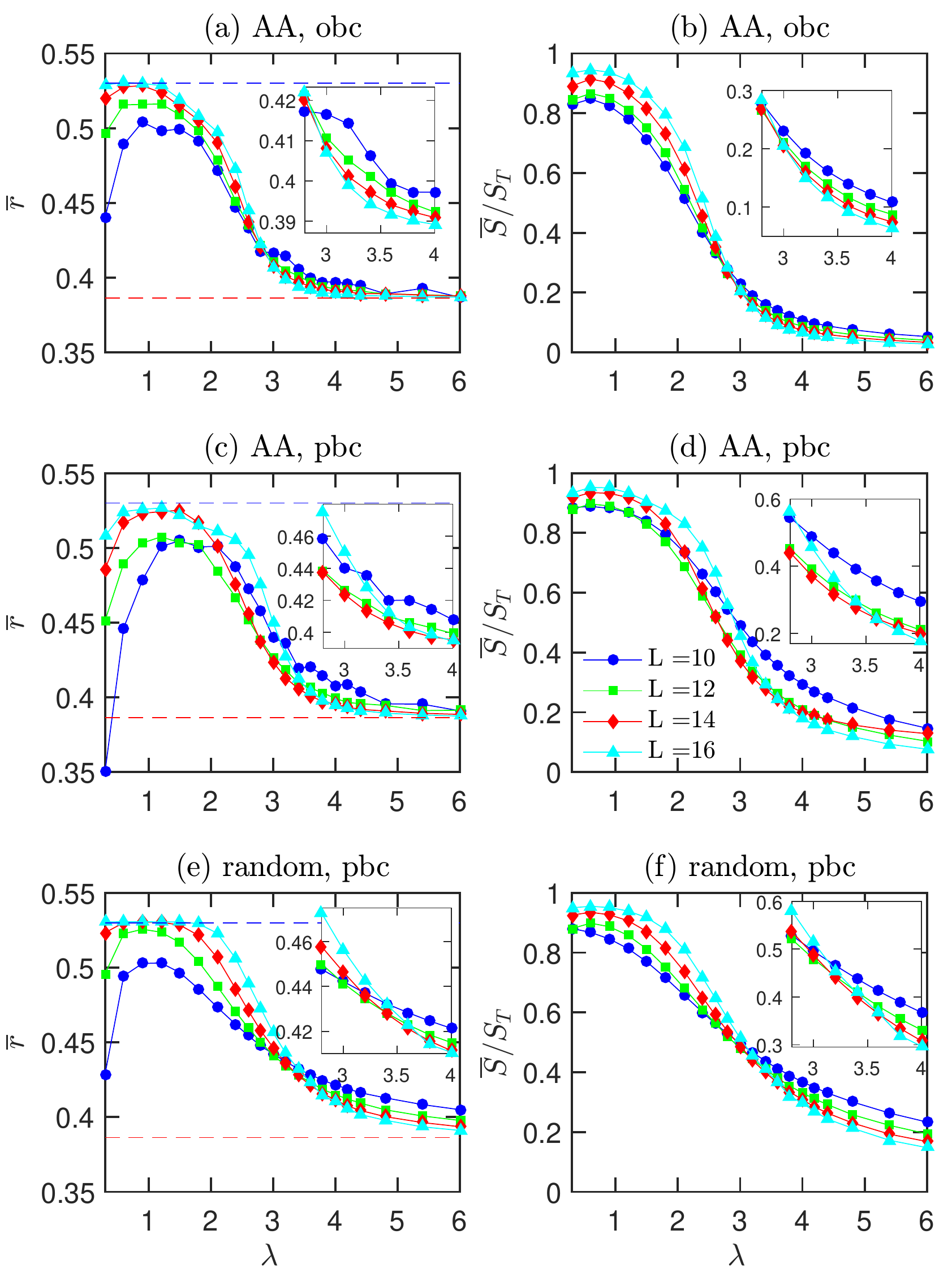}
\end{center}
\caption{Left panel: disorder-averaged adjacent gap ratio $\bar{r}$ vs disorder strength $\lambda$ for different system size $L$ for (a) AA model with an open boundary condition (obc), (c) AA model with a periodic boundary condition (pbc), and (e) random model with a pbc. The blue and red dashed lines denote the values of $\bar{r}$ for Gaussian orthogonal ensemble (GOE) and Poissons distributions, respectively. Right panel: disorder-averaged half-chain entanglement entropy $\bar{S}/S_T$ vs disorder strength $\lambda$ for different system size $L$ for (b) AA model with a obc, (d) AA model with a pbc, and (f) random model with a pbc. All quantities are calculated from the middle $\frac{1}{3}$ of the spectrum. Inset: zoom-in plot near MBL transition.}\label{fig:fig1}
\end{figure}

We determine the critical disorder strength $\lambda_c$ at which the MBL transition happens from the disorder-averaged adjacent gap ratio $\bar{r}$ and half-chain entanglement entropy $\bar{S}$. The level statistics are parametrized by the adjacent gap ratio given by $r_n = \min(\delta_n,\delta_{n+1})/\max(\delta_n,\delta_{n+1})$ where $\delta_n = E_{n+1} - E_{n}$ is the spacing between adjacent energy levels. The entanglement entropy $\bar{S}$ and its standard deviation $\Delta_S$ is divided by the entanglement entropy $S_T = 0.5 [L\mathrm{ln} (2) -1]$ for a pure state drawn randomly~\cite{Page-1993}. The range of the standard deviation $\Delta_S/S_T$ is between 0 and 0.5 as the value of $S/S_T$ is between 0 and 1. 
For the AA model, we compare $\bar{r}$ and $\bar{S}/S_T$ computed using periodic and open boundary conditions as shown in Figs.~\ref{fig:fig1}(a)-\ref{fig:fig1}(d). We take the location of the crossing in $\bar{r}$ and $\bar{S}/S_T$ to estimate the MBL transition. As the crossing is drifting with increasing $L$, we  take the location of the MBL transition to be the crossing between the data for the second largest ($L = 14$) and the largest ($L = 16$) system size that we have. We observe much larger finite-size effects near the crossing of the data for periodic chains as opposed to that for open chains. Since the finite-size effect for the periodic AA model is quite substantial, we take the MBL transition for the AA model from the open boundary condition case (middle panel of Fig.~\ref{fig:fig1}).  For the AA model, the MBL transition is at $\lambda_c \gtrsim 3 t$.
For the random model, the finite-size effects are not as severe with periodic boundary conditions, as seen in Figs.~\ref{fig:fig1}(e) and \ref{fig:fig1}(f), and we therefore take $\lambda_c \gtrsim 3.6t$ for the random model. We find that the transition in the AA model is slightly less than that of the random model which is consistent with Ref.~\cite{Khemani-2017}, but we do take both of them as lower bounds. 

\section{Transport Properties Across the Phase Diagram}~\label{sec:transport}

To probe the existence and size of the subdiffusive regime in either of these models, we consider the disorder-average optical conductivity $\sigma(\omega)$ and return probability $R(\tau)$. In this paper, we focus on the dynamical transport properties of the Hamiltonian in Eq.~(\ref{eqn:ham}) within linear response. We therefore compute the Kubo expression for the optical conductivity at infinite temperature ($T \rightarrow \infty$), which is given by
\begin{equation}\label{eq:cond}
T \sigma(\omega) = \frac{\pi}{Z L}\sum_{n,m}|\langle n |J | m\rangle|^2\delta(\omega-\omega_{nm}).
\end{equation}
Here, $|n \rangle$ are the many-body eigenstates, $J=-it\sum_j(c_{j}^{\dag}c_{j+1} - \mathrm{H.c.})$ is the current operator, $\omega_{nm}=\omega_n - \omega_m$ is the difference between the many-body eigenenergies, and $Z$ is the total number of states. Throughout this paper, we are going to denote $T\sigma(\omega)$ as simply $\sigma(\omega)$. In our numerical calculations, we approximate the $\delta$ function in Eq.~\eqref{eq:cond} by a Lorentzian function with a width $\eta = \delta/10$, where $\delta = \sqrt{L}/2^L$ scales with the average level spacing of the system. We discuss the dependence of our results on the broadening width $\eta$ in Appendix~\ref{sec:broadening}.

To study the dynamics in real time, we also evaluate the return probability $R(\tau) = \langle n_j(\tau)n_j(0) \rangle$ where the averaging is taken over all eigenstates, samples and sites. We observe much larger finite-size effects in the time domain and will clearly state what is an artifact of not reaching a sufficiently large system size to probe the long-time dynamics of the system. 

\subsection{DC conductivity}
\begin{figure}[h!]
\begin{center}
\includegraphics[width=\linewidth]{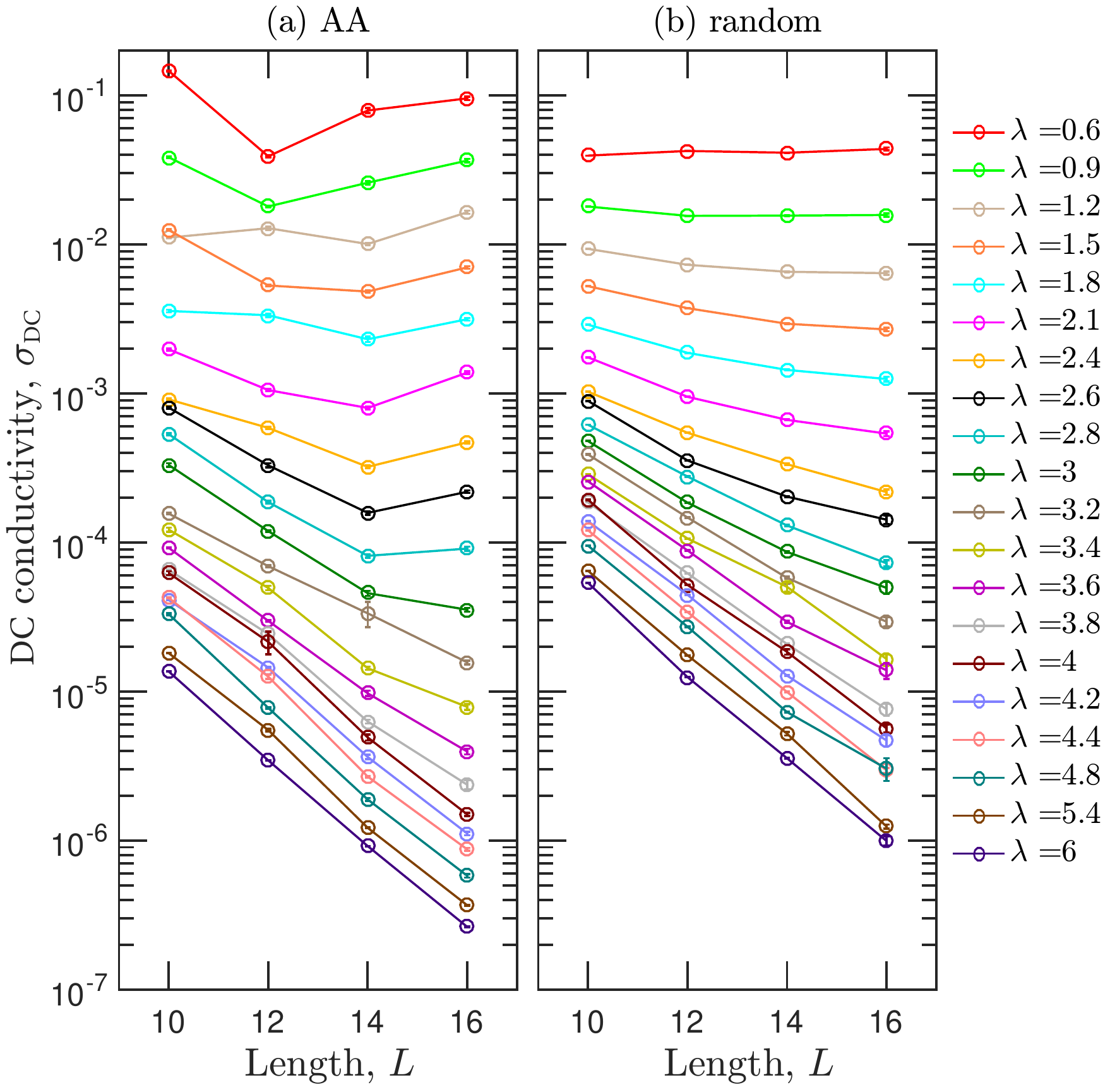}
\end{center}
\caption{Disorder-averaged DC conductivity $\sigma_{\mathrm{DC}}$ vs system size $L$ for different disorder strength $\lambda$ for (a) AA and (b) random models. The DC conductivity is calculated from the middle $\frac{1}{3}$ of the spectrum. 
}\label{fig:fig2}
\end{figure}

We first focus on the DC limit of the conductivity $\sigma_{\mathrm{DC}}\equiv\sigma(\omega=0)$ and consider the system size dependence across the phase diagram, directly comparing AA and random models, as shown in Fig.~\ref{fig:fig2}. Note that we must take periodic boundary conditions to compute $\sigma_{\mathrm{DC}}$, as open boundary conditions always yield $\sigma_{\mathrm{DC}}=0$. 
For consistency, we compute $\sigma(\omega)$ and $R(\tau)$ using periodic boundary conditions in the next subsection.

Let us first focus on the AA model whose $\sigma_{\mathrm{DC}}$ is shown in Fig.~\ref{fig:fig2}(a). At small $\lambda$, we find $\sigma_{\mathrm{DC}}$ is increasing with $L$ (almost) saturating to a constant at $\lambda=0.6t$, signifying a diffusive regime. For increasing $\lambda$ in the thermal regime, the conductivity is decreasing  for each $L$ but we find a finite size effect not present in the random model where $\sigma_{\mathrm{DC}}$ increases with increasing $L$ in going from the second largest system size  ($L=14$) to the largest systems that we have reached ($L=16$). 
Our data are consistent with $\sigma_{\mathrm{DC}}$ vanishing with $L$ near $\lambda \approx \lambda_c$, very close to the MBL transition. Thus, we do not find a \emph{clear} subdiffusive regime in the finite-size scaling of $\sigma_{\mathrm{DC}}$. Now turning to the random model in Fig.~\ref{fig:fig2}(b), we find $\sigma_{\mathrm{DC}}$ is $L$ independent for the largest $L$ and $\lambda \lesssim 1.2t$. For $\lambda$ in the range of $1.2t \lesssim \lambda \lesssim \lambda_c (= 3.6t)$, we find $\sigma_{\mathrm{DC}}$ vanishes like a power law in $1/L$, which is consistent with a subdiffusive regime. For the MBL phase in both models, we find clear insulating behavior $\sigma_{\mathrm{DC}}\sim e^{-aL}$.

\begin{figure*}[t!]
\begin{center}
\includegraphics[width=\linewidth]{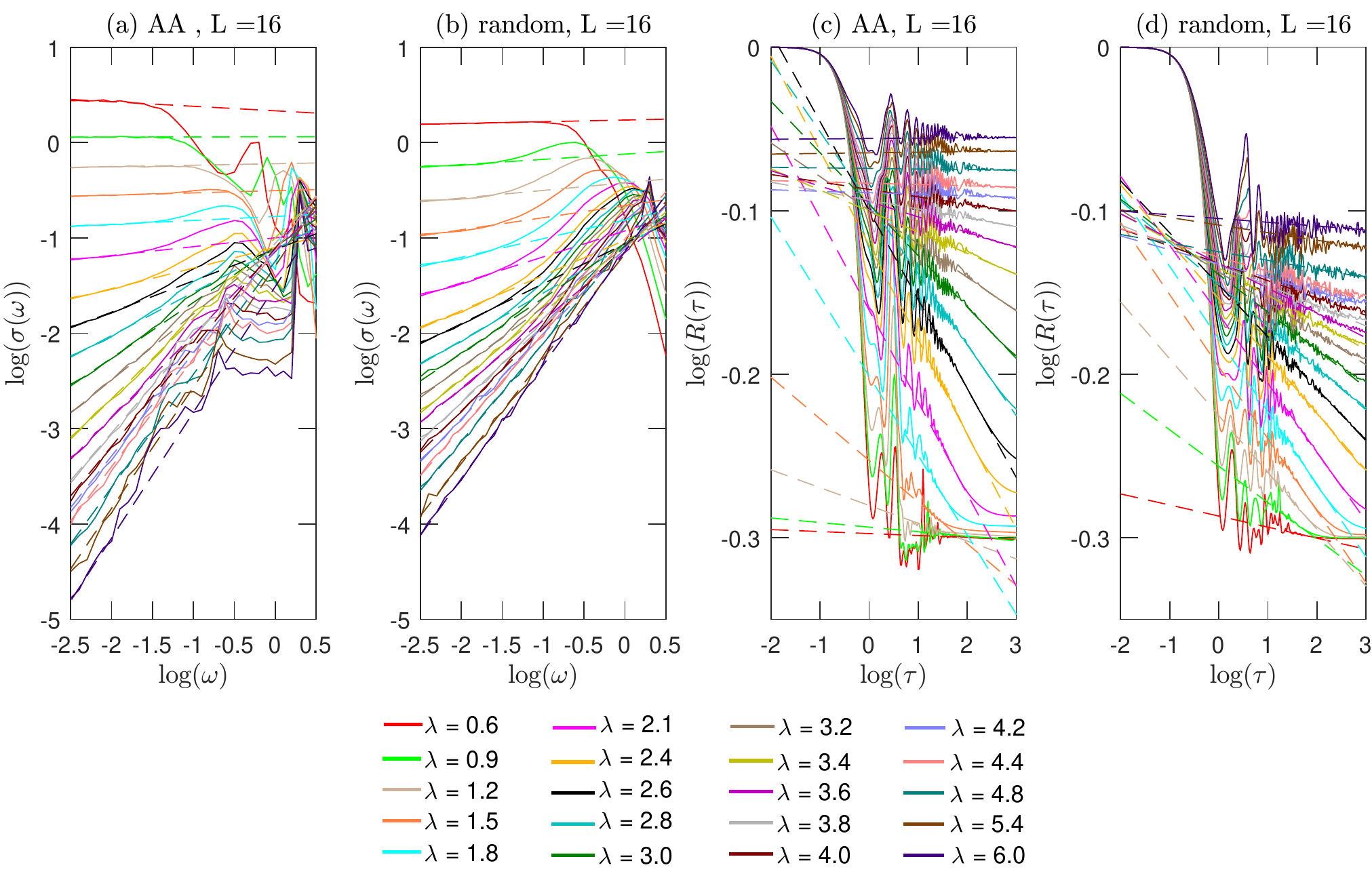}
\end{center}
\caption{Disorder-averaged optical conductivity $\sigma(\omega)$ vs frequency $\omega$ of (a) AA and (b) random models with system size $L = 16$ for different disorder strength $\lambda$. The dashed line is the fit to the equation $\sigma(\omega) \sim\omega^\alpha$. Disorder-averaged return probability $R(\tau)$ vs time $\tau$ of (c) AA and (d) random models with system size $L = 16$ for different disorder strength $\lambda$. The dashed line is the fit to the equation $R \sim \tau^{-\beta}$.  All the quantities are computed from the middle $\frac{1}{3}$ of the spectrum.}\label{fig:fig3}
\end{figure*}

In comparing the AA model with the random model, we find that  in the thermal phase ($\lambda < \lambda_c$), the DC conductivity of the AA model is larger compared to that of the random model. In contrast, in the MBL phase the DC conductivity is much smaller for the AA model than that for the random model. This can naturally be explained by the absence of any rare regions in the AA model. In the AA model, the thermal phase is much more metallic due to the absence of any rare insulating bottlenecks. This leads to a larger average $\sigma_{\mathrm{DC}}$ that does not go to zero at large $L$. On the other hand, in the MBL phase there are no rare thermalizing or metallic bubbles to make the conductivity large. As a result, the AA model is a much better insulator with $\sigma_{\mathrm{DC}}$ almost an order of magnitude smaller than its random counterpart at $\lambda=6t$.

\subsection{Dynamical conductivity and return probability}
We can also probe the existence of the subdiffusive regime by considering the low-energy dynamics across the phase diagram. To this end, we now turn to our results on $\sigma(\omega)$ and $R(\tau)$. We show the data for $L=16$ in Fig.~\ref{fig:fig3} but have also considered $L=12$ and $14$. We fit the low-frequency behavior of $\sigma(\omega)$ and the long-time behavior of $R(\tau)$ to a power-law form, i.e.,
\begin{subequations}
\begin{align}
\sigma(\omega) &\sim \omega^{\alpha},\label{eqn:powerlaw}\\
R(\tau) &\sim 1/\tau^{\beta},
\end{align}
\end{subequations}
where the exponents are related to the dynamic exponent $z$ in the thermal phase via $\alpha=1-2/z$ and $\beta=1/z$ \cite{2015_Agarwal_MBL_PRL}. Thus, for $\alpha=0$ the behavior is diffusive ($z=2$) and for $\alpha>0$, it is subdiffusive ($z>2$).

\begin{figure}[t!]
\begin{center}
\includegraphics[width=\linewidth]{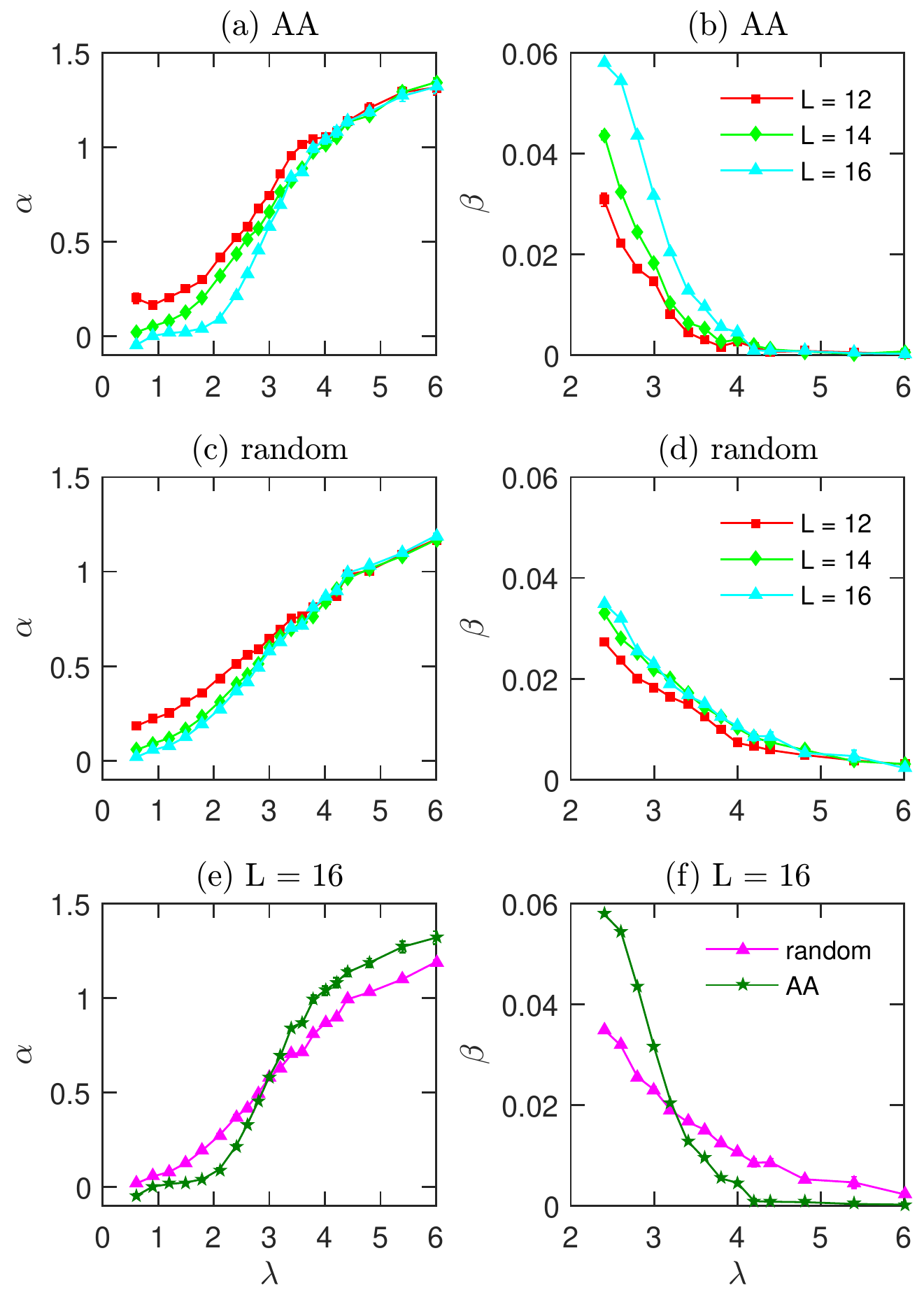}
\end{center}
\caption{Top panel: The exponent (a) $\alpha$ of $\sigma(\omega) \sim \omega^\alpha$ and (b) $\beta$ of the return probability $R \sim \tau^{-\beta}$ for the AA model with different system sizes. Middle panel: The exponent (c) $\alpha$ and (d) $\beta$ for the random model with different system sizes. Bottom panel: Comparison between (e) $\alpha$ and (f) $\beta$ for AA and random models for $L = 16$.}\label{fig:fig4}
\end{figure}

Our data for $\sigma(\omega)$ serve as our best estimate of the diffusive and subdiffusive regime [see Figs.~\ref{fig:fig3}(a) and~\ref{fig:fig3}(b)]. 
For small $\lambda$, we find $\sigma(\omega)$ is relatively flat in $\omega$ for both models. We find the diffusive regime  extends all the way to $\lambda \lesssim  2.1t$ for the AA model, whereas in the random model it ends at $\lambda \lesssim 1.2 t$. Thus, our results for the dynamical conductivity show 
that the diffusive regime within the thermal phase  is more stable in the quasiperiodic model compared to the random case. In addition, we find
that there is a subdiffusive regime in the AA model for these \emph{finite-size} systems. 
Our estimate for the size of the subdiffusive regime in the random model from $\sigma(\omega)$ is consistent with the estimate from $\sigma_{\mathrm{DC}}$.  

We now turn to the return probability $R(\tau)$ in the long-time limit [see Figs.~\ref{fig:fig3}(c) and~\ref{fig:fig3}(d)]. For small $\lambda$, it is difficult to reach the asymptotic long-time limit to probe the diffusive length scale in the problem. For example, in the regime where $\sigma_{\mathrm{DC}}$ is $L$ independent in the random model (which implies $z=2$ and $\beta=1/2$), there is almost no power-law regime in $\tau$ of our return probability data. This ``flat'' large-$\tau$ behavior deep in the thermal regime is an artifact of our finite-size numerics not having access to long enough time scales to probe the asymptotic diffusive scaling regime. We find this flat finite-size-limited regime is larger in the AA model. For $\lambda\geq 2.4t$, we have a large enough system size to begin to probe the asymptotic scaling regime. Entering the MBL phase, we find $\beta \rightarrow 0$ and the long-time behavior is essentially flat, consistent with previous studies for the random model~\cite{2015-Torres-Dynamics,2017-Torres-Extended}.  Comparing random and AA models, we find that $R(\tau\rightarrow \infty)$ is significantly larger in the AA model than in the random model. This implies that the memory of an initial state is retained much better in a quasiperiodic MBL phase and is consistent with the quasiperiodic MBL phase being more robust against delocalization than its random counterpart. 

Our results for $\alpha$ and $\beta$ are shown in Fig.~\ref{fig:fig4}. 
For small $\lambda$ deep in the thermal phase, the finite-size corrections are significant in the return probability due to not having access to the long-time behavior. Therefore, we only show $\beta$ for $\lambda \ge 2.4t$.
Since the diffusive regime has $\alpha = 0$ ($\beta=1/2$) and deep in the MBL phase $\alpha \rightarrow 2$ ($\beta\rightarrow 0$) (see Ref.~\cite{2015_Gopalakrishnan_PRB}), the exponent $\alpha$ ($\beta$) has to increase (decrease) for increasing $\lambda$ as the model passes through the critical regime near $\lambda_c$.
The system size dependence of the extracted exponents is shown in Figs.~\ref{fig:fig4}(a)--\ref{fig:fig4}(d). In the AA model we find that $\alpha(L)$ [$\beta(L)$] is a decreasing (increasing) function of increasing $L$ in the thermal regime; the subdiffusive regime [defined by $\alpha(L)>0$], shrinks with increasing $L$. In contrast, our estimate of $\alpha$ ($\beta$) in the random model has a much weaker $L$ dependence, displaying an essentially $L$-independent broad subdiffusive regime. 

The direct comparison of $\alpha$ for the AA and random models is shown in Fig.~\ref{fig:fig4}(e). 
We find that near the transition but in the thermal phase $\alpha_{\mathrm{random}}>\alpha_{\mathrm{AA}}$ for $\lambda < \lambda_c^{\mathrm{AA}}$, consistent with insulating rare bottlenecks requiring a finite $\omega$ to activate transport. On the other hand, on the MBL side near the transition, this is reversed, i.e., $\alpha_{\mathrm{random}}<\alpha_{\mathrm{AA}}$, where the lack of any rare thermal region in the AA model makes it harder to thermalize (and hence less metallic). In addition, we find the slope of $\alpha$ versus $\lambda$ is steeper in the AA model. All of this is consistent with an $L$-dependent (small) subdiffusive regime in the AA model where $\alpha(L)$ is displaying a sharp crossover.     

In the MBL phase, the exponentially vanishing $\sigma_{\mathrm{DC}}$  with system size $L$ implies that $z = \infty$. Near $\lambda_c$ but on the MBL side of the transition,  finite-size systems create a  crossover regime to a diverging dynamic exponent at $\lambda^*$. We can estimate the location of this crossover boundary by the point where $\beta\rightarrow 0$. As shown in Figs.~\ref{fig:fig4}--\ref{fig:fig4}(f), we find that this crossover boundary (at the largest $L=16$) occurs at $\lambda^* \approx 4t$ and $\lambda^* \gtrsim 6t$ for the AA and random models, respectively. 
This provides further evidence that the MBL phase in the AA model is more stable than in the random system as $\beta\approx 0$ survives down to a smaller value of the potential strength.

\section{Sample-to-sample and state-to-state fluctuations}\label{sec:sigmadcdist}
\subsection{Distributions of $\sigma_{\mathrm{DC}}$}

\begin{figure}[h]
\begin{center}
\includegraphics[width=\linewidth]{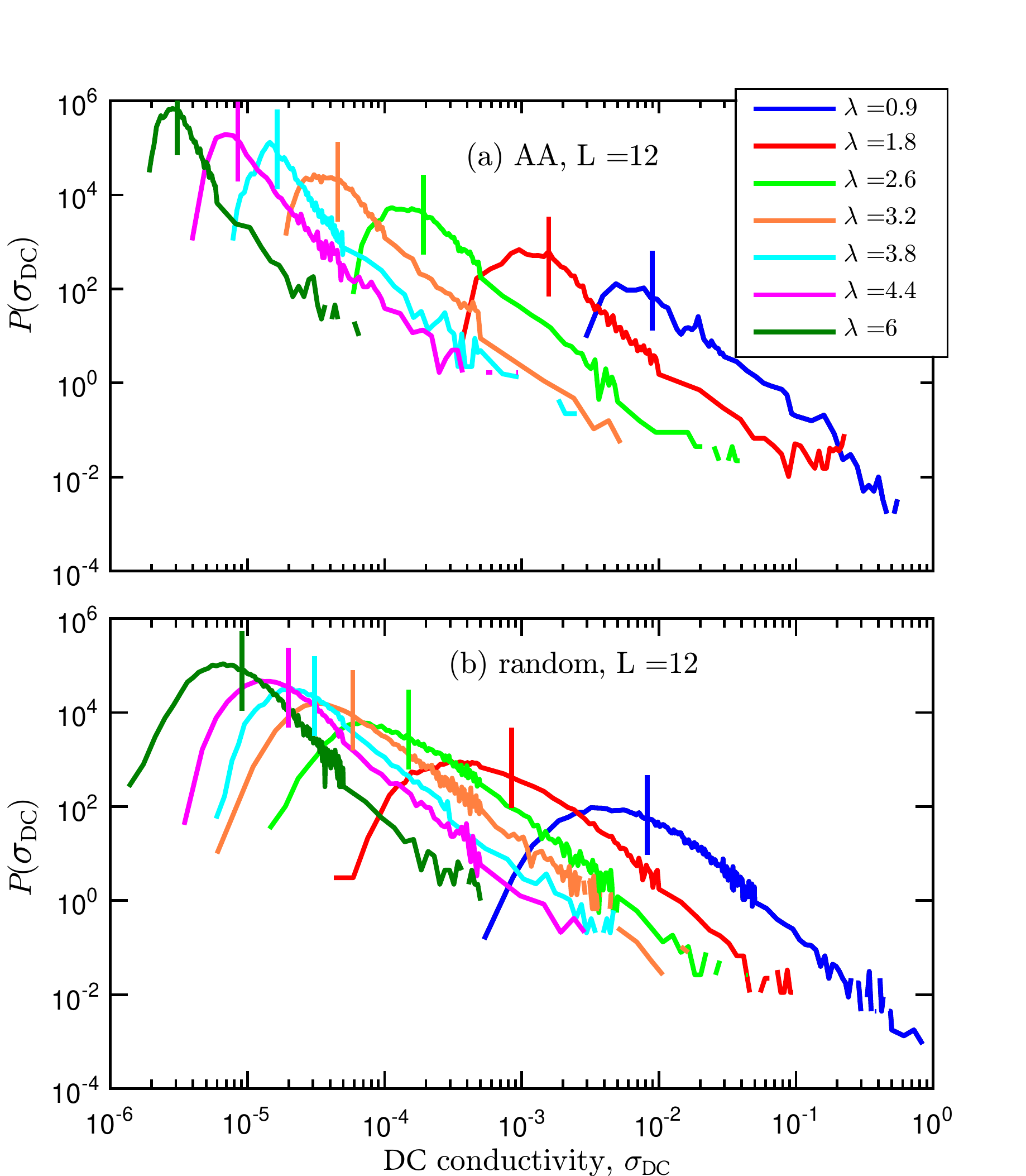}
\end{center}
\caption{Probability density functions of DC conductivity $P(\sigma_{\mathrm{DC}})$ of AA (top panel) and random (bottom panel) models for $L = 12$ and different disorder strength $\lambda$. 
The vertical solid lines denote the median of the distributions. }\label{fig:fig8}
\end{figure}

To access sample-to-sample fluctuations of the DC conductivity, we now consider the distribution of the DC conductivity $P(\sigma_{\mathrm{DC}})$ for a large number of disorder realizations (=20,000). We first consider the distributions for a fixed $L=12$ across the phase diagram, directly comparing AA and random models as shown in Fig.~\ref{fig:fig8}. Some features in $P(\sigma_{\mathrm{DC}})$ are common to both models: For weak potential strengths deep in the thermal phase we find $P(\sigma_{\mathrm{DC}})$ is characterized by a peak near the median with a power-law tail towards large $\sigma_{\mathrm{DC}}$. Surprisingly, we find that the large-conductivity power-law tail survives across the phase diagram but is falling off faster in the MBL phase.

\begin{figure}[h]
\begin{center}
\includegraphics[width=\linewidth]{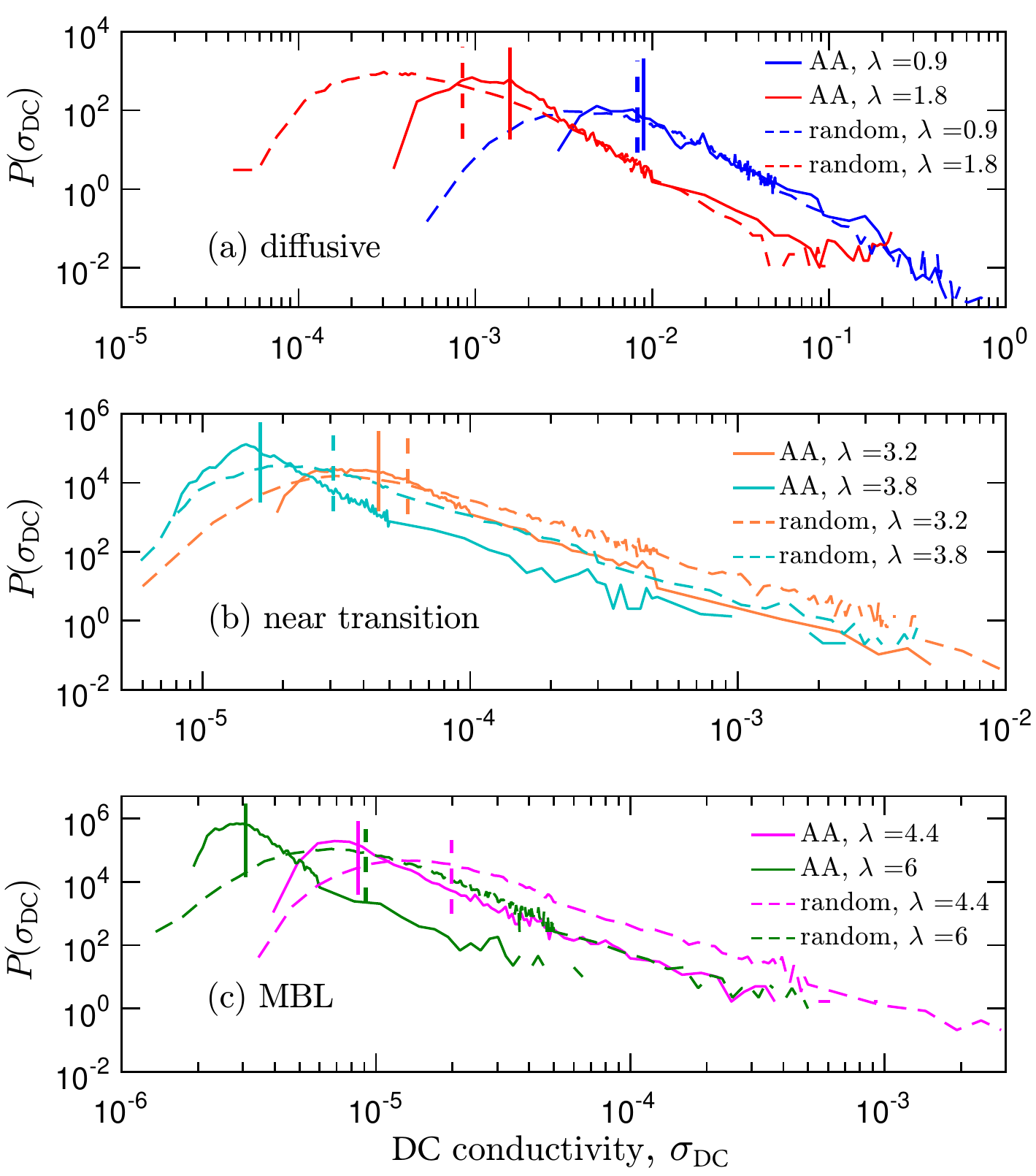}
\end{center}

\caption{Comparison between the probability density functions of DC conductivity $P(\sigma_{\mathrm{DC}})$ of AA (solid line) and random (dashed line) models for $L = 12$ and different disorder strength $\lambda$. The vertical solid and dashed lines denote the corresponding median value of the distributions. The DC conductivity is calculated from the middle $\frac{1}{3}$ of the spectrum.}\label{fig:fig6}
\end{figure}

\begin{figure}[h]
\begin{center}
\includegraphics[width=\linewidth]{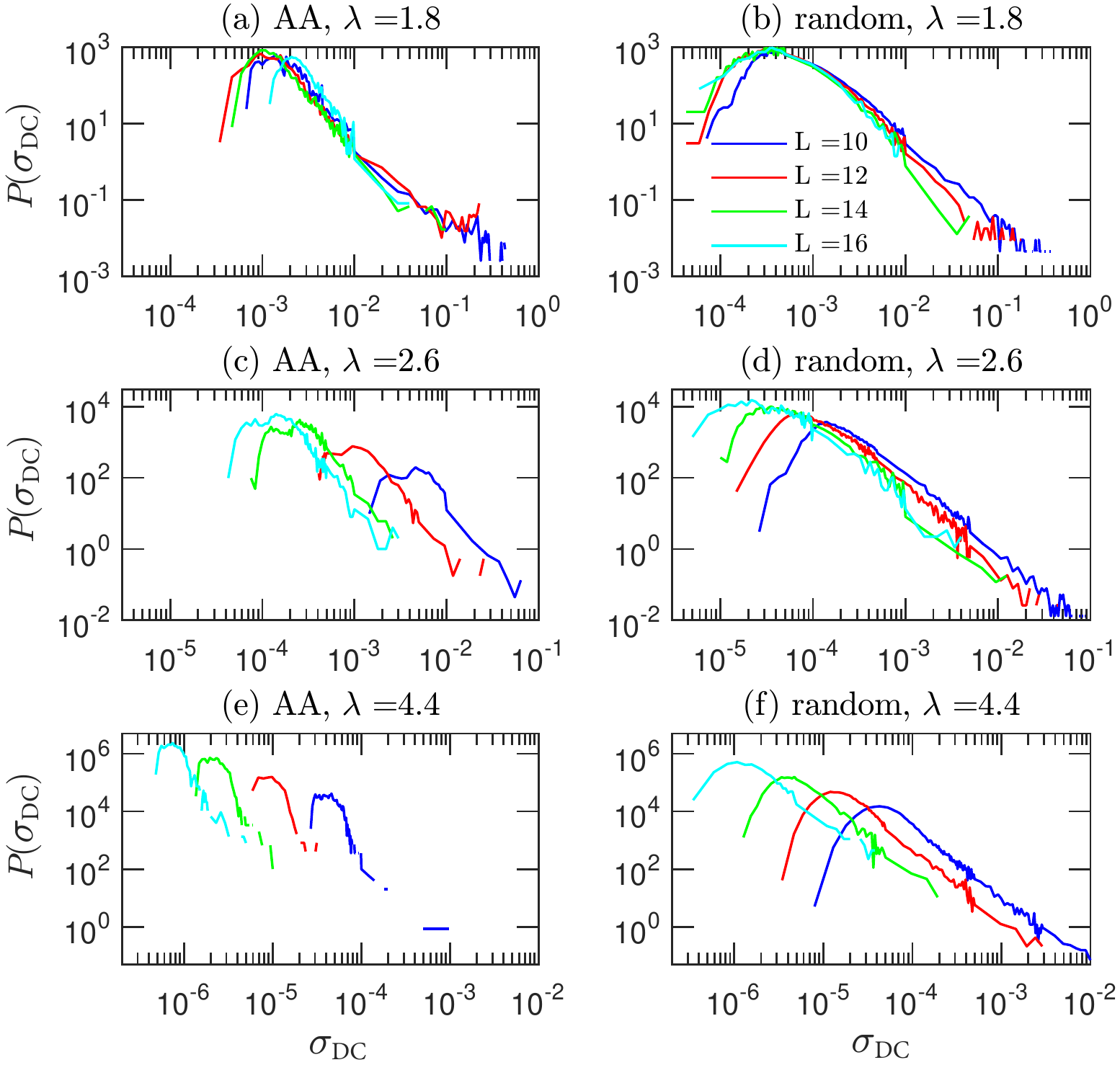}
\end{center}
\caption{Probability density functions of DC conductivity $P(\sigma_{\mathrm{DC}})$ of AA (left panel) and random (right panel) models for different system size $L$ and disorder strength $\lambda$ in diffusive (top panel), near the transition (middle panel), and MBL (bottom panel) regions. 
}\label{fig:fig7}
\end{figure}

Interestingly, the distributions are also very different across the two models. We make a direct comparison of random and AA models for each regime of the model for $L=12$ in  Fig.~\ref{fig:fig6}.  In the thermal phase, the large-conductivity tails essentially match between the random and AA models but we find that the width of the peak  in $P(\sigma_{\mathrm{DC}})$ extends to much smaller values of $\sigma_{\mathrm{DC}}$ in the random model as compared to the AA case. A common feature in the AA data is that the peak looks like it is ``cut off'' at small $\sigma_{\mathrm{DC}}$. As shown in Fig.~\ref{fig:fig6}(b), near the transition we find the random case has a broad distribution extending past both the maximum and minimum $\sigma_{\mathrm{DC}}$ in the AA model at the same $\lambda$. In addition, the power-law tail near the transition has more statistical weight in the random model. In the MBL phase of the AA model, the width of the peak is narrow, approximately one order of magnitude smaller than that of the random case for $\lambda=6t$. Also, in the MBL phase we find the large-conductivity tail extends to substantially larger values in the random model as opposed to that of AA, being separated by about one order of magnitude. Lastly, in the MBL regime we find that the peak is reasonably well fit by a log-normal distribution for the random model only (not shown), but this does not capture the tail towards large $\sigma_{\mathrm{DC}}$. 

Now that we have a feeling for the behavior of $P(\sigma_{\mathrm{DC}})$ across the phase diagram, we consider the system size dependence of $P(\sigma_{\mathrm{DC}})$ in each regime of the model in Fig.~\ref{fig:fig7}. In the thermal regime with $\lambda=1.8t$, we find $P(\sigma_{\mathrm{DC}})$ has a weak $L$ dependence near the peak and spreads out for large $\sigma_{\mathrm{DC}}$ for the random model. For the AA model, we find the power-law tail is roughly $L$ independent but the peak is sharpening  up with increasing $L$. The decrease of the peak width with increasing $L$ is consistent with diffusive samples. In the finite-size subdiffusive regime, we find the peak extends to increasingly smaller $\sigma_{\mathrm{DC}}$ with increasing $L$ and the power-law tail remains pronounced in the random model whereas in the AA model, the power-law tail is strongly suppressed. In the MBL phase, $P(\sigma_{\mathrm{DC}})$ is markedly different between the two models. In the AA model, the peak is very narrow with a width that is decreasing with increasing $L$ and has a weak power-law tail. On the contrary, the random model has a well-defined peak which is an order of magnitude broader than the AA case for each $L$.

Our results in this section on $P(\sigma_{\mathrm{DC}})$ have established another clear distinction between random and quasiperiodic interacting many-body systems.
It is natural to associate the broadness of the distributions in the random model to rare Griffith effects: In the thermal phase, the rare samples that produce local insulating bottlenecks contribute to the small $\sigma_{\mathrm{DC}}$  part of the peak. On the other hand, in the MBL phase, local thermal regions contribute to the large-conductivity power-law tail, both of these features are very pronounced in the random model relative to the AA data. Near the transition, the random model has contributions from both of these types of samples, which leads to a broad distribution. However, for the thermal phase in the AA model, we find the peaks sharpen up with increasing $L$ and there is an $L$-independent power-law tail.  In the MBL phase of the AA model, we expect that the dominant excitations are  MBL-Mott resonances~\cite{2015_Gopalakrishnan_PRB}, which  contribute to the power-law tail. We expect that the size of the MBL-Mott resonances grow as we approach the transition from the MBL side, which give weight to the power-law tail towards large $\sigma_{\mathrm{DC}}$. On the other hand, the states that contribute to the peak in the MBL phase are insulating and not (or weakly) resonant. 

\begin{figure}[h]
\begin{center}
\includegraphics[width=\linewidth]{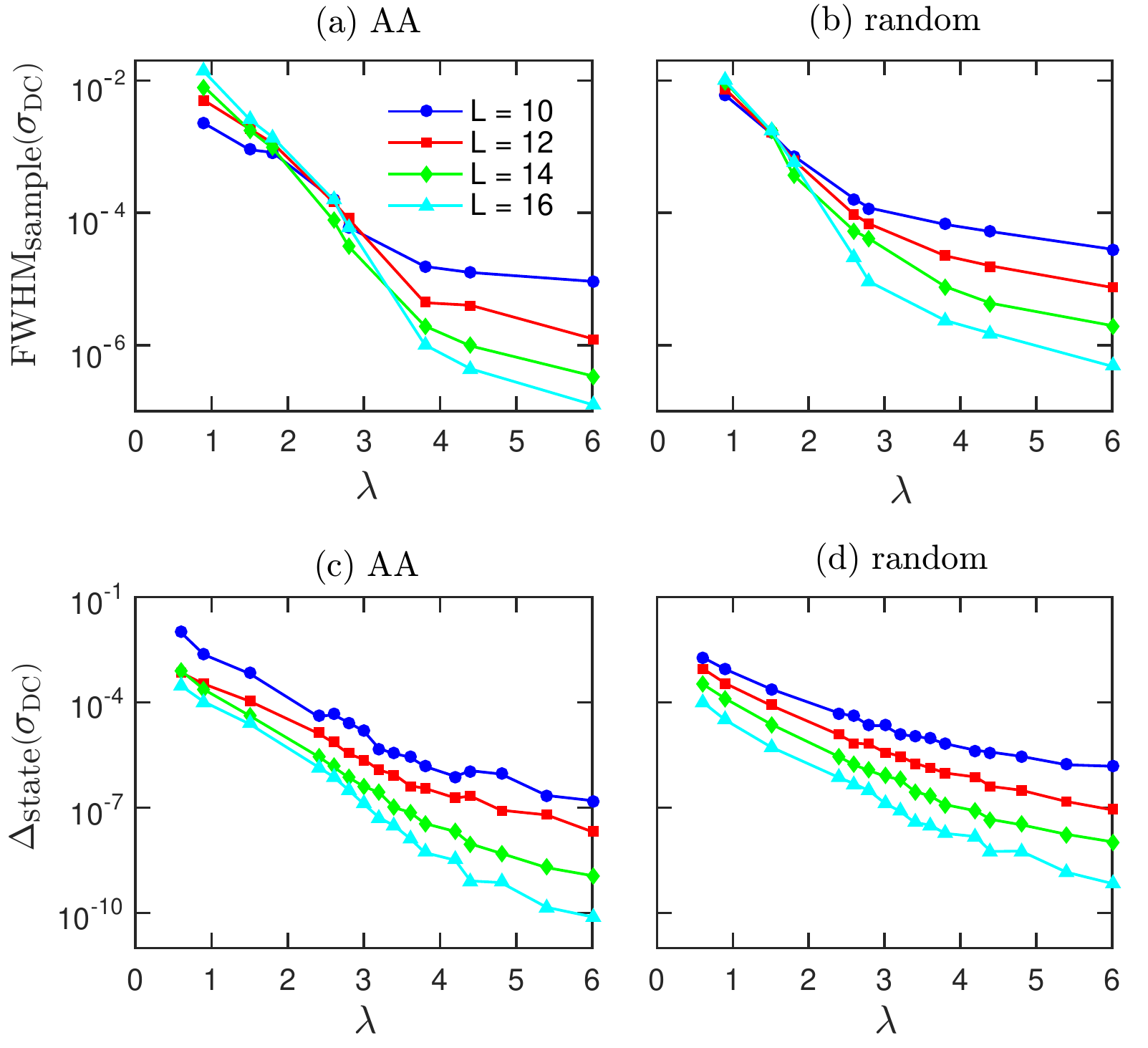}
\end{center}
\caption{Full width at half maximum over samples of the probability distribution of $\sigma_{\mathrm{DC}}$   [$\mathrm{FWHM}_{\mathrm{sample}}(\sigma_{\mathrm{DC}})$] vs disorder strength $\lambda$ for (a) AA and (b) random models.
Standard deviation of $\sigma_{\mathrm{DC}}$ over eigenstates [$\Delta_{\mathrm{state}}(\sigma_{\mathrm{DC}})$] vs disorder strength $\lambda$ for (c) AA and (d) random models. 
}\label{fig:FWHM_sigmadc}
\end{figure}

\subsection{Sample-to-sample and state-to-state contributions to $\sigma_{\mathrm{DC}}$}
We  quantify the sample-to-sample fluctuations of $\sigma_{\mathrm{DC}}$ by studying the width of the peak in $P(\sigma_{\mathrm{DC}})$ by computing its full width at half-maximum (FWHM), as shown in Figs.~\ref{fig:FWHM_sigmadc}(a) and~\ref{fig:FWHM_sigmadc}(b). 
Common to both the AA and random models, we find that in the diffusive regime of the thermal phase the FWHM increases with increasing $L$, whereas in the MBL regime the FWHM decreases with increasing $L$.
Interestingly, for the AA model we find that the FWHM has a crossing (that is drifting with $L$) near the MBL transition while for the random model, the crossing is very weakly $L$ dependent and occurs near the entrance to the subdiffusive regime ($\lambda \approx 1.2t$). Thus the sample-to-sample contributions in random and AA models are markedly distinct. Note that we choose to show the FWHM here instead of the standard deviation of $\sigma_{\mathrm{DC}}$ over samples [calculated from Eq.~\eqref{eq:cond} by averaging $\sigma_{\mathrm{DC}}$ across samples] as the standard deviation over samples is noisy due to 
requiring a very large number of samples to accurately capture
the long large-conductivity tail of the $P(\sigma_{\mathrm{DC}})$.

It is important to contrast this measure of  sample-to-sample variations with fluctuations across eigenstates. To do so, we also compute the standard deviation of $\sigma_{\mathrm{DC}}$ over eigenstates [$\Delta_{\mathrm{state}}(\sigma_{\mathrm{DC}})$]. This quantity is obtained by first summing over $m$ in Eq.~\eqref{eq:cond} and taking the standard deviation over the index $n$ and  then averaging across samples. As shown in Figs. ~\ref{fig:FWHM_sigmadc}(c) and (d),  $\Delta_{\mathrm{state}}(\sigma_{\mathrm{DC}})$ are qualitatively similar between the AA and random models: they both decrease with increasing $\lambda$ and system size $L$. We find the dependence on $L$ of $\Delta_{\mathrm{state}}(\sigma_{\mathrm{DC}})$ is stronger in the MBL phase in both AA and random models. However, we do find quantitative distinctions between the AA and random models, i.e., in the thermal regime $\Delta_{\mathrm{state}}(\sigma_{\mathrm{DC}})$ is larger in the AA model  and in the MBL regime this is reversed with $\Delta_{\mathrm{state}}(\sigma_{\mathrm{DC}})$ being an order of magnitude smaller in the AA model. 

\subsection{Sample-to-sample contributions to $\sigma(\omega)$ and entanglement entropy}
We now connect the sample-to-sample fluctuations in transport to that in the entanglement.
We quantify the effect of sample-to-sample fluctuations on $\sigma(\omega)$ by parsing our data based on small and large values of $\sigma_{\mathrm{DC}}$. To be consistent across the phase diagram, we separately average the data that have either $\sigma_{\mathrm{DC}}$ greater than or less than the median of $P(\sigma_{\mathrm{DC}})$. We call these ``large'' and ``small'' conductivity contributions. This is a natural separation as the states that make up the peaks and tails of the conductivity distributions of the random model are dominated by Griffith contributions of different character.  After averaging over small and large $\sigma_{\mathrm{DC}}$ separately, we compute the low-frequency scaling regime following Eq.~(\ref{eqn:powerlaw}), which allows us to define the power laws $\alpha_{\mathrm{small}}$ and $\alpha_{\mathrm{large}}$ that reflect the power law after averaging over small ($\sigma_{\mathrm{DC}} < \sigma_{\mathrm{DC}}^{\mathrm{median}}$) and large ($\sigma_{\mathrm{DC}} > \sigma_{\mathrm{DC}}^{\mathrm{median}}$) DC conductivity, respectively.  The difference between the two exponents:
\begin{equation}
\Delta \alpha = \alpha_{\mathrm{small }}-\alpha_{\mathrm{large}}
\end{equation}
directly probes the spread of sample contributions to $\sigma(\omega)$. We normalize this quantity by its (expected) maximum variation $(=2$) and plot this quantity for $L = 16$ in Fig.~\ref{fig:Dalpha} while the fits for $\alpha_{\mathrm{small}}$ and $\alpha_{\mathrm{large}}$ are shown in Appendix~\ref{sec:deltaalpha}. In the random model, since Griffith effects  dominate at the MBL transition, where both rare insulating blocks and rare thermal bubbles contribute, $\Delta \alpha$ is peaked near $\lambda_c$. Thus, $\Delta \alpha$ measures  the spread of different types of samples and therefore, in the random model its magnitude measures the strength of Griffith effects. In contrast, $\Delta \alpha$ in the AA case has a broad peak centered around $\lambda_c$ and has a much smaller magnitude compared to its random counterpart. Interestingly, we find that $\Delta \alpha$ rises from zero at the same value of disorder strength that marks the subdiffusive regime from  $\alpha$ and $\beta$  for $L=16$ in the AA model (cf. Figs.~\ref{fig:fig4} and ~\ref{fig:Dalpha}).

\begin{figure}[h]
\begin{center}
\includegraphics[width=\linewidth]{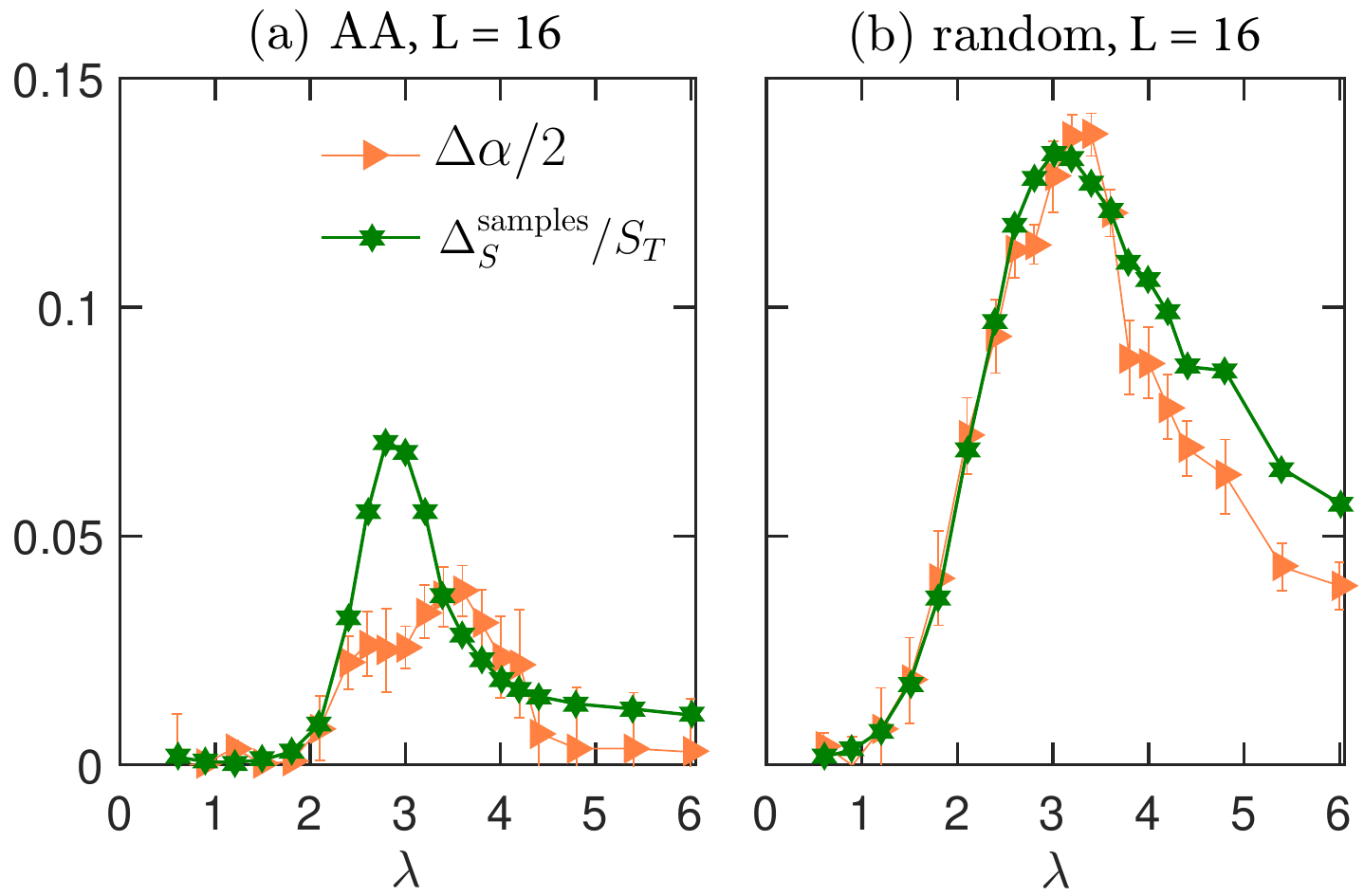}
\end{center}
\caption{Plots of $\Delta \alpha$ and $\Delta_S^{\mathrm{samples}}$ vs $\lambda$ for (a) AA and (b) random models. $\Delta \alpha$ = $\alpha_{\mathrm{small}} - \alpha_{\mathrm{large}}$ is the difference between the power-law exponent $\alpha$ of sample-averaged $\sigma(\omega)$ having $\sigma_{\mathrm{DC}} < \sigma_{\mathrm{DC}}^{\mathrm{median}}$ ($\alpha_{\mathrm{small}}$) and $\sigma_{\mathrm{DC}} > \sigma_{\mathrm{DC}}^{\mathrm{median}}$ ($\alpha_{\mathrm{large}}$) with $\sigma_{\mathrm{DC}}^{\mathrm{median}}$ being the median value of the DC conductivity. We compare this $\Delta \alpha$ with the standard deviation of half-chain entanglement entropy over different samples ($\Delta_{S}^{\mathrm{samples}}$) and find that the region of $\lambda$ over which the entanglement entropy and transport are dominated by the sample-to-sample variations are similar. The value of $\Delta \alpha$ is normalized by 2, which is the difference between $\alpha$ for the pure thermal and pure MBL region, while the value of $\Delta_{S}^{\mathrm{samples}}$ is normalized by $S_T = 0.5[L \mathrm{ln}(2) -1]$, which is the entanglement entropy of a randomly drawn pure state.
}\label{fig:Dalpha}
\end{figure}

We now compare the signatures of sample-to-sample  in the transport to that in the entanglement entropy.
To understand the different and relevant contributions, we parse our data following Refs.~\cite{Khemani-2016,Khemani-2017} and calculate the standard deviation of half-chain entanglement entropy over samples, $\Delta^{\mathrm{samples}}_S$, which is computed by first taking the average of the half-chain entanglement entropy over eigenstates and spatial cuts in a given sample and then computing the standard deviation of the averaged entanglement entropy over different samples. We have also considered fluctuations over samples and cuts and our results and system size dependence (as shown in Appendix~\ref{sec:stdsent}) are similar to the results in Ref.~\cite{Khemani-2016,Khemani-2017} on slightly different models. We show $\Delta^{\mathrm{samples}}_S/S_T$ for $L=16$ and compare this with $\Delta \alpha/2$ in Fig.~\ref{fig:Dalpha}. While the overall magnitude is sensitive to the choice of normalization, it is rather striking that the trends in both transport and entanglement are so similar where  the rise and fall of the peak (near $\lambda_c$) agree. Thus, we conclude that the sample-to-sample fluctuation dominated regimes in transport and entanglement are similar. In addition, we find that fluctuations over samples in both transport and entanglement are much larger in the random system compared to the AA model.

\section{Transport near the thermal-to-MBL transition}\label{sec:scenario}
\subsection{Activated dynamical scaling near the MBL transition}
\begin{figure}[t!]
\begin{center}
\includegraphics[width=\linewidth]{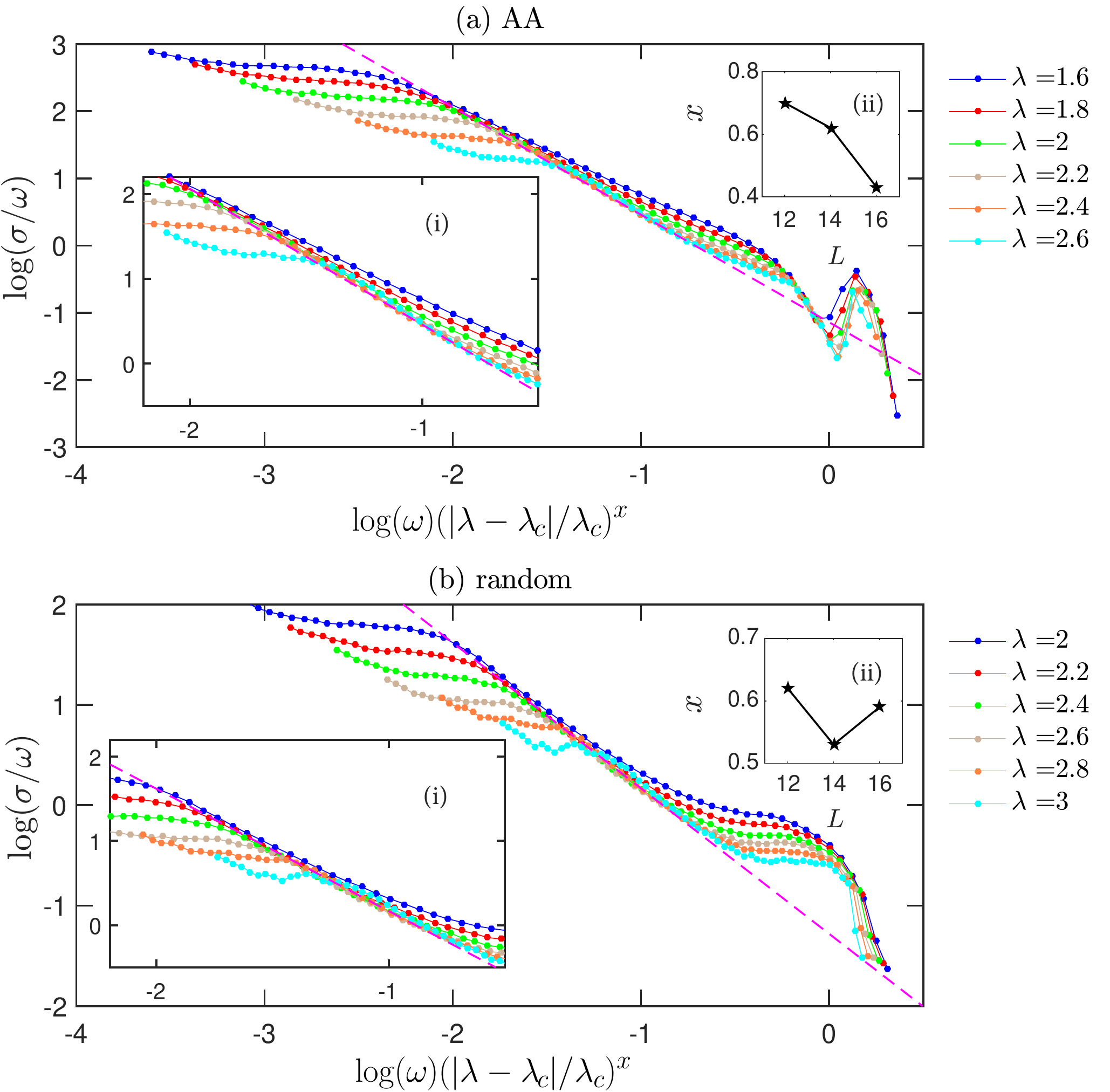}
\end{center}
\caption{Activated dynamical scaling of log[$\sigma(\omega)/\omega$]) vs log($\omega$)($|\lambda-\lambda_c|/\lambda_c)^{x}$ for (a) AA and (b) random models with $L=16$. The conductivity $\sigma(\omega)$ is calculated  over the middle $\frac{1}{3}$ of the spectrum. The critical value $\lambda_c$ at which the MBL transition happens is taken to be 3$t$ and 3.6$t$ for the AA and random models, respectively.  We find the best scaling collapse occurs for $x \approx 0.4$ and $x \approx 0.6$ for AA and random models, respectively. The dashed line is a guide to the eye for the data collapse. Note that the data for the AA model do not collapse well using the ansatz in Eq.~\eqref{eq:actdynscaling}. Inset (i): zoom-in version of the data collapse. We also collapsed the data for $L = 14$ and 12 (not shown). For the AA model, we obtained the best data collapse exponent $x$ to be $\approx 0.6$ and $\approx 0.7$ for $L =14$ and $12$, respectively, while for the random model, we obtained $x\approx 0.53$ and $0.62$ for $L =14$ and $12$, respectively. As shown in inset (ii), the value of $x$ for the AA model decreases with increasing $L$ while the value of $x$ for the random model is roughly independent of $L$. Since only the data for the random model collapse well, we take $x = \nu\psi$ and use the estimate of $\nu \approx 0.7$ for the random model (see Appendix~\ref{sec:datacollapse}) to get the value of $\psi$. We find the value of $\psi \approx 0.9$ for the random model.}\label{fig:fig11}
\end{figure}

We now come to the implications of our results on the nature of transport near the thermal-to-MBL transition. We aim to ask whether the dynamic exponent $z$ diverges like a power law on approach to the MBL transition similar to the random model. 
In order to probe the scaling regime associated with a dynamic exponent diverging near the MBL transition, we use the activated dynamical scaling ansatz to compare the AA and random critical regimes. If $z$ is continuously varying and diverges in a power-law fashion on approach to the MBL transition  from the thermal side (consistent with strong disorder RG treatments \cite{2015_Potter_PRX, 2015_Vosk_PRX,Parameswaran-2016}), then $z\sim \xi^{\psi}$, where $\psi$ is the activated dynamical scaling exponent and $\xi$ is the correlation length that diverges like $\xi\sim (|\lambda-\lambda_c|/\lambda_c)^{-\nu}$ with the correlation length exponent $\nu$. Applying this scaling ansatz to the dynamical conductivity in the thermal phase $\sigma(\omega) \sim \omega^{1-2/z}$, we find
\begin{eqnarray}\label{eq:actdynscaling}
\log(\sigma/\omega) \sim  \log(\omega)\left(\frac{|\lambda-\lambda_c|}{\lambda_c}\right)^{x},
\end{eqnarray}
 where $x$ can be regarded as a fit parameter and we can associate $x=\nu \psi$ if the data do collapse. As shown in Fig.~\ref{fig:fig11} for $L=16$ our data for the random model displays a clear scaling regime in the argument $Y\equiv\log(\omega)(|\lambda-\lambda_c|/\lambda_c)^{x}$ for slightly over a decade, with an exponent $x$ that depends weakly on $L$. At small $Y$, our data falls off this scaling form, which we have checked is due to the finite-size cutting off of the low-frequency behavior (not shown). At larger $Y$, the data splays out systematically for decreasing $\lambda$. Taking $x=\nu\psi$ and using our estimate of $\nu$ on these small system sizes ($\nu \approx 0.7$ for the random model as shown in Appendix~\ref{sec:datacollapse}), we find  $\psi\approx0.9$ for the random model, which is rather close to the expectation from the strong disorder RG, which finds $\psi=1$ ~\cite{2015_Potter_PRX, 2015_Vosk_PRX,Parameswaran-2016}. This suggests that in the random model $\psi$ does not suffer from as large finite-size effects as compared with $\nu$. 
 
 In contrast, for the AA case we do not find a clear scaling regime: the attempted collapse of the data spreads out for increasing $Y$, and where the small region of collapse breaks down, the data splays out more strongly than the random case for smaller values of the scaling argument $Y$ [see insets (i) of Fig.~\ref{fig:fig11}]. Moreover, the extracted fit exponent $x$ depends more strongly on $L$  than the random model [see insets (ii) of Fig.~\ref{fig:fig11}]. These results imply that our ansatz that $z$ diverges like a power law (in the small-$\omega$ limit) for the AA model seems not to  apply. Thus, the failure of the activated dynamical scaling and the fact that $\alpha = 1 - 2/z$ (for $\lambda < \lambda_c$) approaches a step function at the transition as $L$ increases [as shown in Fig.~\ref{fig:fig4}(a)] suggest that $z$  
 jumps
 from $2$ to $\infty$  in the large-$L$ limit 
 across the MBL transition in quasiperiodic systems.

\subsection{Finite-size crossover diagram}
\begin{figure}[h!]
\begin{center}
\includegraphics[width=0.8\linewidth]{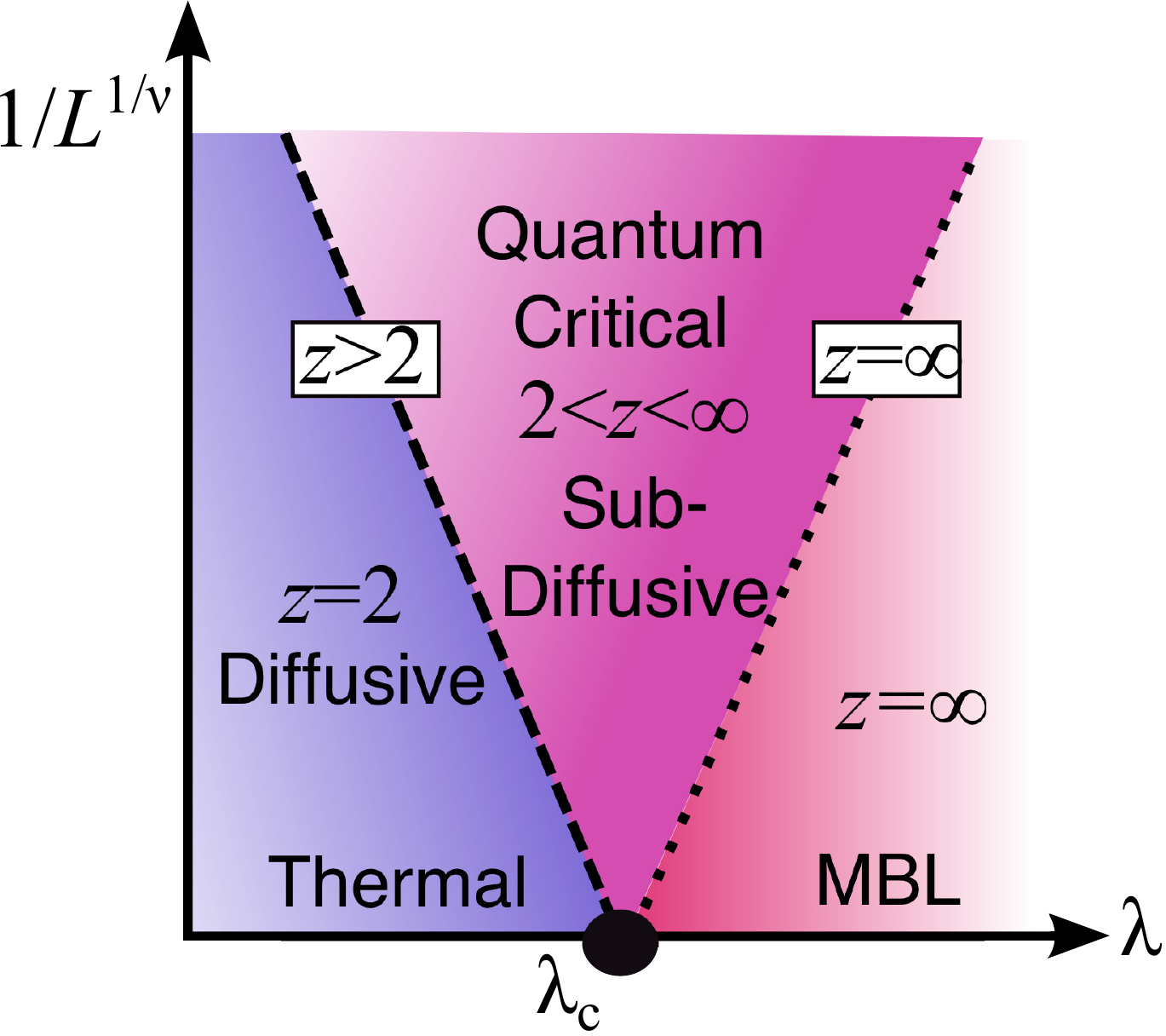}
\end{center}
\caption{Schematic crossover diagram for quasiperiodic systems with a finite system size $L$ versus the incommensurate potential strength $\lambda$. The vertical axis is plotted as $1/L^{1/\nu}$ (where $\nu$ is the correlation length exponent) so that the crossover lines separating each respective regime that are dictated by the correlation length $\xi$ are straight lines. We have included the dynamic exponent $z$ across the crossover diagram. Based on our numerical data, we argue that the finite-size quantum critical crossover regime is subdiffusive. In the thermodynamic limit, our data are consistent with a transition from a thermal diffusive phase to an MBL phase, different from the case of random systems. 
 }\label{fig:pd}
\end{figure}

With all of our numerical results on the transport properties in hand, we are in a good position to develop a physical scenario for the origin of the subdiffusive regime in quasiperiodic systems. We must reconcile the appearance of a subdiffusive regime with the failure of activated dynamical scaling. To do so, it is instructive to think of our data in terms of a finite-size crossover diagram ($1/L^{1/\nu}$ on the vertical axis and $\lambda$ on the horizontal axis), where the correlation length $\xi \sim (|\lambda - \lambda_c|/\lambda_c)^{-\nu}$ dictates the finite-size crossover boundaries to the thermal and MBL regimes on either side of $\lambda_c$ at finite $L$ (see Fig.~\ref{fig:pd}). Emanating from the thermal-to-MBL transition is a quantum critical fan that is anchored by the transition at infinite $L$ and is dictated by the universal scaling properties set by $\xi$. 
Such a construction for the MBL transition of the random model has been given in Ref.~\cite{Khemani-2016}, which clearly finds the crossover boundary in the local entanglement entropy (with the subsystem containing a single site) for $\lambda < \lambda_c$ but  the crossover from the quantum critical to MBL regime is more subtle. 

In the random model, fluctuations in the vicinity of the critical point come from the combination of the diverging length scale associated with the transition and separate Griffith contributions. It is very hard to disentangle these two effects, and both of them will contribute to observables such as the width of the peak in  $\Delta^{\mathrm{samples}}_S$ and $\Delta \alpha$ in Fig.~\ref{fig:Dalpha}(b). However,
in the quasiperiodic model, since there are no Griffith effects and as we have shown sample-to-sample fluctuations are much weaker, we expect that fluctuations near $\lambda_c$ come from critical eigenstates with a diverging length scale on these system sizes.   We can therefore take the location of the crossover boundary from the thermal to quantum critical regime to be where $\Delta^{\mathrm{samples}}_S$ starts to grow as $\lambda$ increases. The boundary for this \emph{matches} the subdiffusive crossover boundary obtained from our transport data. 

We therefore argue that  critical eigenstates  give rise to the subdiffusive transport regime we have observed on finite-size systems. Correlated sample-to-sample fluctuations will produce different values of the correlation length $\xi$ for each state (that are all on the order of $L$) and these fluctuations can produce the peak in $\Delta \alpha$ in Fig.~\ref{fig:Dalpha} (a). This leads us to construct the schematic finite-size crossover diagram for the quasiperiodic MBL transition in Fig.~\ref{fig:pd}, where the quantum critical regime is subdiffusive. The crossover out of the quantum critical regime to the MBL regime is much more subtle, but we can roughly take it to be marked by 
where $\beta\approx 0$. Consistent with our numerical data for the quasiperiodic model in Figs.~\ref{fig:fig4} and ~\ref{fig:fig11}, which show that at the transition, $\alpha$ approaches a step function as the system size increases and the failure of the activated dynamical scaling, respectively, we expect that  $z$ jumps from $2$ to $\infty$ 
across 
the quasiperiodic transition in the thermodynamic limit. For $\lambda < \lambda_c$ on large system sizes much bigger than $\xi$, the eigenstates will not be critical and  we  therefore argue that the thermal phase is diffusive in quasiperiodic systems in the thermodynamic limit and the MBL transition coincides with the vanishing of the DC conductivity. This makes the quasiperiodic transition fundamentally distinct from that of its random counterpart. 

\section{Discussion and Conclusion}\label{sec:conclusion}

We have theoretically studied the transport properties in both 1D interacting quasiperiodic and random systems at infinite temperature across the MBL transition. We systematically compared and contrasted the dynamical conductivity and the return probability near the MBL transition in quasiperiodic and random models and found a major distinction between them in each phase of the model. Our choice of the quasiperiodic model and its random generalization has allowed us to directly compare both models at the same potential strength. The detailed comparison between the two models has allowed us to unambiguously remove Griffith effects from the problem. We have found this leads to very different system size dependence on the observed subdiffusive regime. 
 We determined the state-to-state and sample-to-sample fluctuations of the conductivity and showed that eigenstate fluctuations are similar in random and quasiperiodic systems, but the random problem is more dominated by  fluctuations across different random samples. Moreover, the distributions of the DC conductivity in the quasiperiodic model have much weaker tails and cover a smaller range than those in the random system.

Our results  have established that both the diffusive regime and the MBL phase in quasiperiodic systems are much more robust than their random counterparts.  The robustness of the quasiperiodic MBL phase is exemplified by both the DC conductivity and the long-time limit of the return probability: $\sigma_{\mathrm{DC}}$ is an order of magnitude smaller than its random counterpart and the memory of the initial state is retained significantly better in the quasiperiodic system. We expect the robustness of the quasiperiodic MBL phase will persist to higher dimensions, where there are no non-perturbative Griffith effects available to locally thermalize the system that have the potential to destabilize the MBL phase altogether. We note that our findings for the stability of the MBL phase obtained from the dynamical transport properties agree nicely with the static entanglement entropy study in Ref.~\cite{Khemani-2017}. Furthermore, our work establishes that the MBL transition in the quasiperiodic model has a finite-size subdiffusive crossover regime which vanishes in the thermodynamic limit as discussed below.

We have shown that the random model obeys activated dynamical scaling with a numerically computed activation exponent $\psi \approx 0.9$. Despite the fact that our estimate of the correlation length exponent $\nu$ strongly violates the CCFS criteria, our estimate of $\psi$ is close to the expectation of the RG. On the other hand, in the quasiperiodic system we have shown this scaling ansatz works poorly on the available system sizes as the data splays out strongly. It will be interesting to try and construct the appropriate scaling ansatz to capture the behavior of the dynamic exponent in quasiperiodic systems. We further argued that the thermal phase in interacting quasiperiodic systems at infinite temperature remains diffusive in the thermodynamic limit and the subdiffusive  transport observed on finite-size systems is due to critical eigenstates inducing a crossover regime.

Our results are consistent with the recent experimental observation~\cite{Luschen2016evidence} of a subdiffusive regime in a cold atom setup of the interacting AA model. We also note that a recent transport study~\cite{lev2017transport} using tDMRG and functional RG also finds a subdiffusive regime in the interacting quasiperiodic model. Due to the finite bond dimension, that acts like a finite-size (or finite-entanglement) effect at quantum critical points~\cite{Pollmann-2009}; these findings are also compatible with our conclusions. Based on our numerical data, we constructed a finite-size crossover diagram for transport in quasiperiodic systems near the MBL transition where the quantum critical crossover regime is subdiffusive and the MBL transition coincides with a metal-to-insulator transition.  Thus, our work establishes  fundamental differences between the thermal and MBL phases as well as the thermal-to-MBL transition in quasiperiodic  and random systems.

\section*{ACKNOWLEDGMENTS}
We thank S. Das Sarma, D. Huse, X. Li, and S. Ganeshan  for useful discussions. We also thank S. Das Sarma, D. Huse, and S. Gopalakrishnan for comments on a draft. This work has been partially supported by JQI-NSF-PFC, LPS-MPO-CMTC, and the PFC seed grant ``Thermalization and its breakdown in isolated quantum systems." We acknowledge the University of Maryland supercomputing resources (http://www.it.umd.edu/hpcc) made available in conducting the research reported in this paper.

\appendix

\section{Effect of level broadening on the conductivity}\label{sec:broadening}
\begin{figure}[h]
\begin{center}
\includegraphics[width=\linewidth]{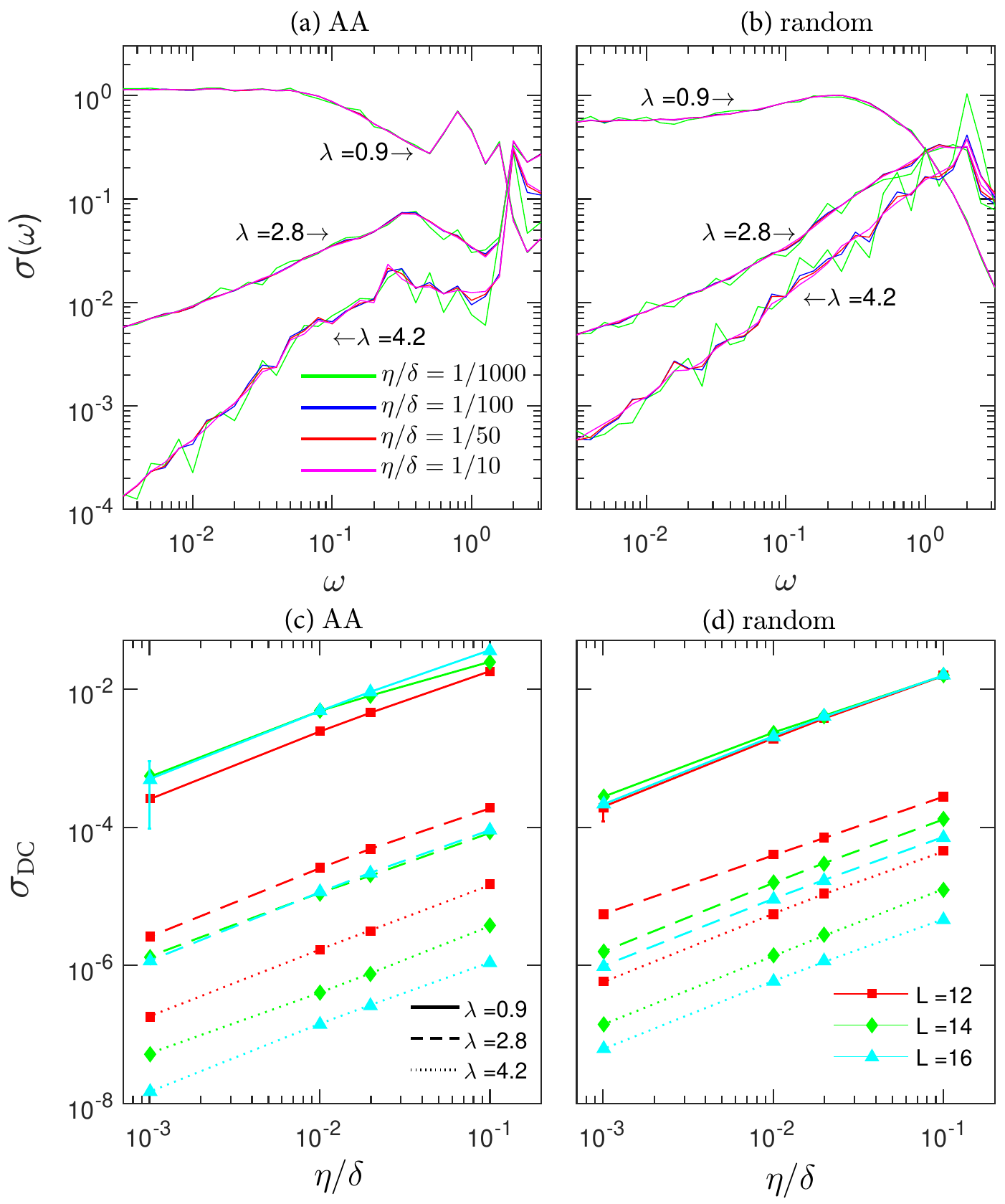}
\end{center}
\caption{Top panel: plot of optical conductivity $\sigma(\omega)$ vs $\omega$ of (a) AA and (b) random models with system size $L = 16$ for various width $\eta$ of the Lorentzian function approximating the delta function in the Kubo formula [Eq.~\eqref{eq:cond}]. The conductivity $\sigma(\omega)$ is calculated from $\frac{1}{3}$ of the full spectrum closest to zero energy. Bottom panel: DC conductivity ($\sigma_{\mathrm{DC}}$) vs $\eta$ of (c) AA and (d) random models for various values of disorder strength $\lambda$ and system size $L$. }\label{fig:fig12}
\end{figure}

In this paper, the conductivity is calculated by using the Kubo formula [Eq.~\eqref{eq:cond}] where we use a Lorentzian function with a width $\eta$ to approximate the delta function. The width $\eta$ is chosen to be smaller than the average level spacing $\delta$ where $\delta = \sqrt{L}/2^L$. The top panel of Fig.~\ref{fig:fig12} shows the plot of $\sigma(\omega)$ vs $\omega$ for different values of $\eta$. For smaller $\eta$, more random realizations are required to reach convergence. We can see from the figure that for the number of realizations used in this paper, our choice of $\eta = \delta/10$ has already reached a convergence. The bottom panel of Fig.~\ref{fig:fig12} shows the dependence of DC conductivity ($\sigma_{\mathrm{DC}}$) on $\eta$ for various system size $L$. As can be seen from the figure, $\sigma_{\mathrm{DC}}$ scales as $\eta$ (Refs.~\cite{Thouless-1981-Conductivity,Berkelbach-2010}). In the MBL phase of both AA and random models, $\sigma_{\mathrm{DC}}$ decreases with increasing $L$.

\section{Variation of conductivity}\label{sec:deltaalpha}
\begin{figure}[h]
\begin{center}
\includegraphics[width=\linewidth]{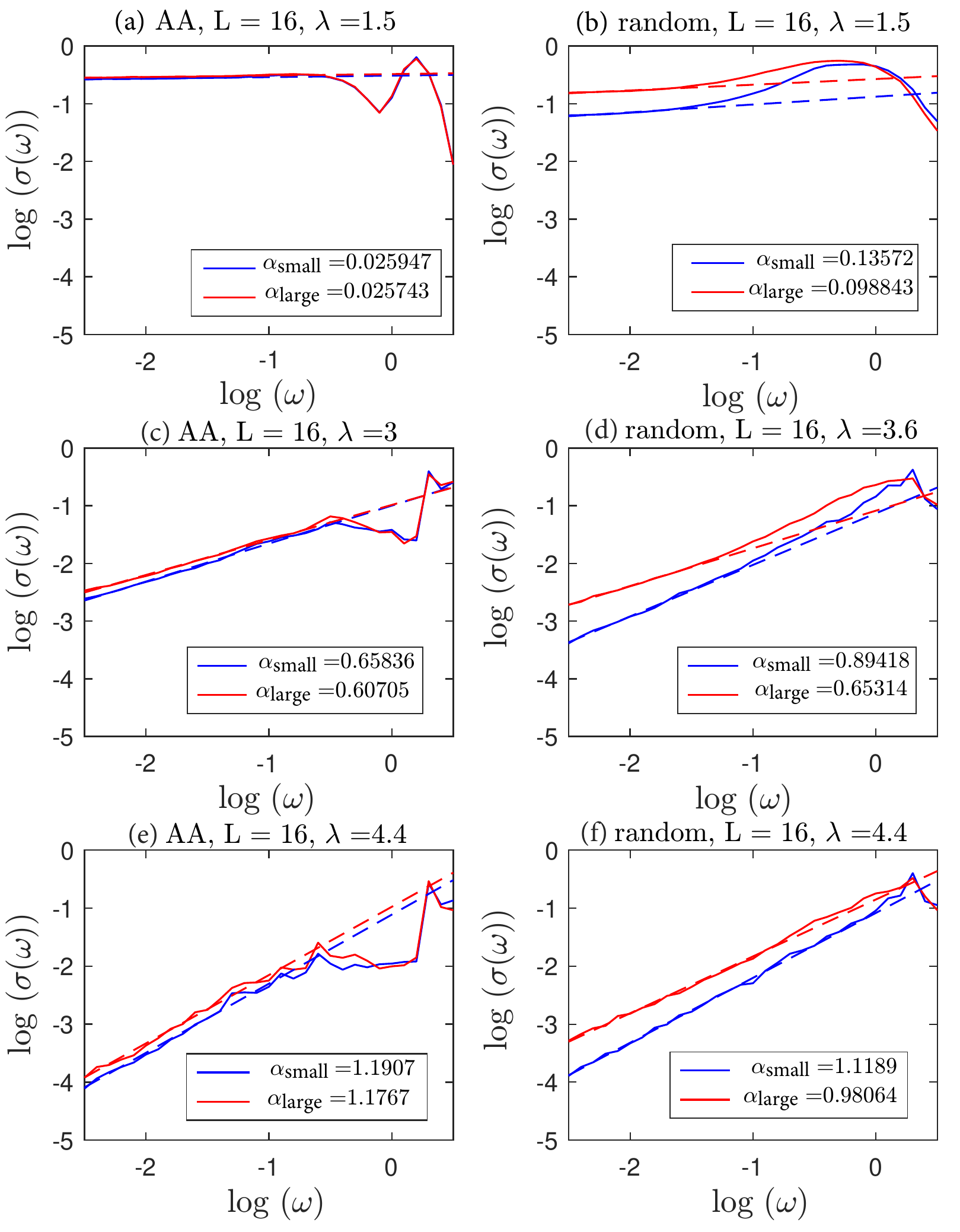}
\end{center}
\caption{Top panel: plot of optical conductivity $\sigma(\omega)$ vs $\omega$ of AA (left panel) and random (right panel) models with system size $L = 16$ for different regions: thermal (top), near transition (middle), and MBL (bottom). The blue (red) curve corresponds to the optical conductivity averaged over samples with $\sigma_{\mathrm{DC}} < \sigma^{\mathrm{median}}_{\mathrm{DC}}$ ($\sigma_{\mathrm{DC}} > \sigma^{\mathrm{median}}_{\mathrm{DC}}$) where $\sigma^{\mathrm{median}}_{\mathrm{DC}}$ is the median value of the DC conductivity. The low-frequency tail is fitted to the formula $\sigma(\omega) \sim \omega^\alpha$ (dashed line) where $\alpha_{\mathrm{small}}$ ($\alpha_{\mathrm{large}}$) corresponds to the  power-law exponent $\alpha$ of $\sigma(\omega)$ with $\sigma_{\mathrm{DC}} < \sigma_{\mathrm{DC}}^\mathrm{median}$ ($\sigma_{\mathrm{DC}} > \sigma_{\mathrm{DC}}^\mathrm{median}$). The conductivity $\sigma(\omega)$ is calculated over $\frac{1}{3}$ of the full spectrum closest to zero energy.}\label{fig:alpha1_metallic_insulating_13}
\end{figure}

In this appendix, we show the fits for the low-frequency power law of  $\sigma({\omega})$ vs $\omega$ used in obtaining Fig.~\ref{fig:Dalpha}. Figure~\ref{fig:alpha1_metallic_insulating_13} shows $\sigma({\omega})$ averaged over samples with $\sigma_{\mathrm{DC}} < \sigma^{\mathrm{median}}_{\mathrm{DC}}$ (blue curve) and $\sigma_{\mathrm{DC}} > \sigma^{\mathrm{median}}_{\mathrm{DC}}$ (red curve) for different regimes: thermal (top), near transition (middle) and MBL (bottom). The low-frequency tail of $\sigma(\omega)$ follows the power law $\sigma(\omega) \sim \omega^\alpha$ as shown by the dashed lines. The difference between the optical conductivity power-law exponent $\Delta \alpha = \alpha_{\mathrm{small}} - \alpha_{\mathrm{large}}$, where $\alpha_{\mathrm{small}}$ and $\alpha_{\mathrm{large}}$ are the exponents of $\sigma(\omega)$ with small ($\sigma_{\mathrm{DC}} < \sigma^{\mathrm{median}}_{\mathrm{DC}}$) and large $\sigma_{\mathrm{DC}}$ ($\sigma_{\mathrm{DC}} > \sigma^{\mathrm{median}}_{\mathrm{DC}}$), respectively, is peaked near the transition and is larger for the random model due to Griffiths effect (as shown in Fig.~\ref{fig:Dalpha}). 

\section{Finite-size critical scaling collapse}\label{sec:datacollapse}
\begin{figure}[h]
\begin{center}
\includegraphics[width=\linewidth]{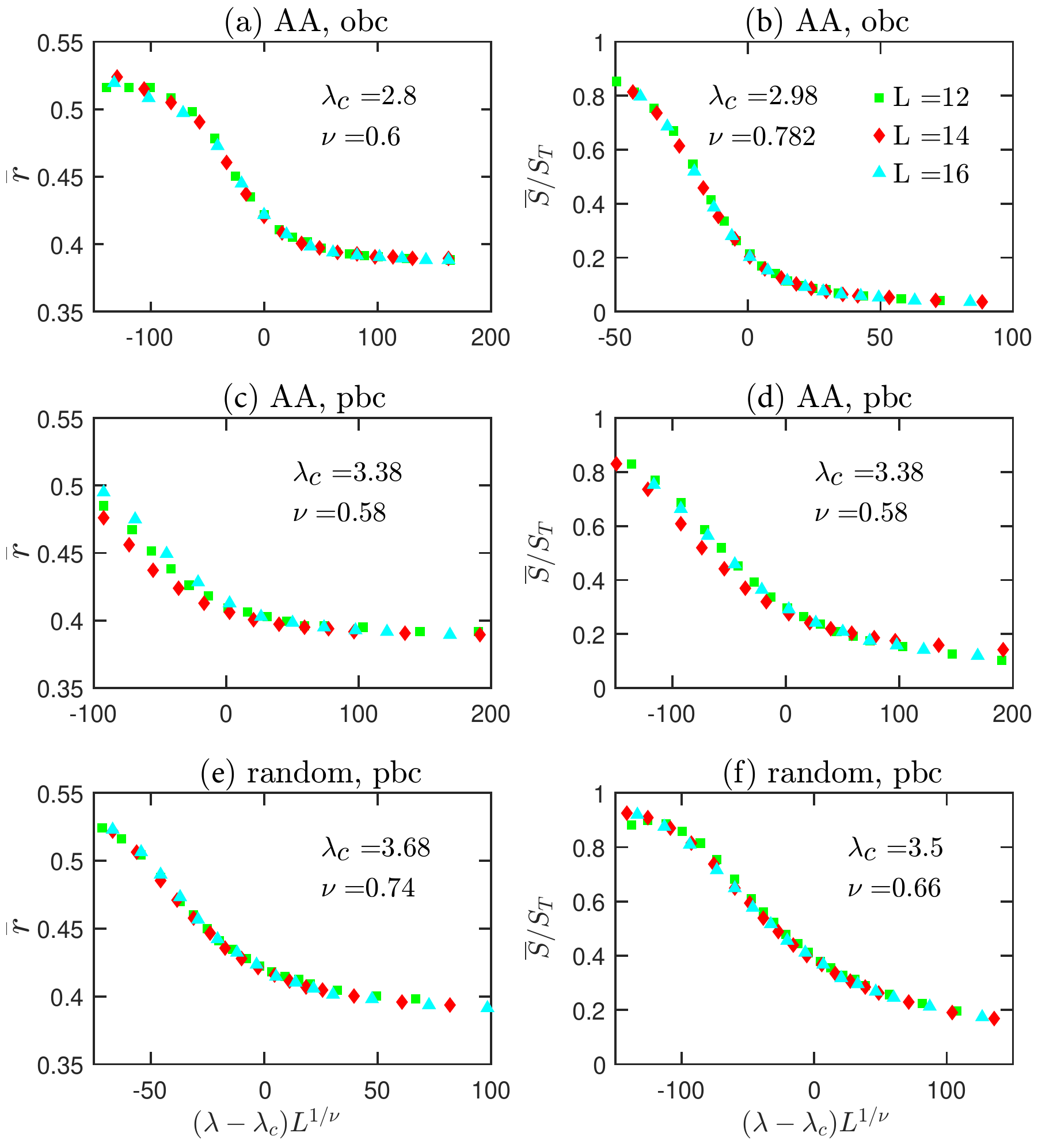}
\end{center}
\caption{Finite-size critical scaling collapse for the level statistics (left panel) and entanglement entropy (right panel) for the AA model with obc (top panel), AA model with pbc (middle panel) and random model with pbc (bottom panel). The value of $\nu$ is $\sim 0.6$ for AA and $\sim 0.7$ for random model. The critical disorder strength $\lambda_c$ at which the MBL transition happens is larger for the random model.}\label{fig:datacollapse}
\end{figure}

In this appendix, we show the finite-size critical scaling collapse for the level statistics and entanglement entropy in Fig.~\ref{fig:datacollapse}. We plot both of these quantities as a function of $(\lambda - \lambda_c)L^{1/\nu}$ where $\nu$ is the correlation length exponent. We find $\nu$ to be $\sim 0.6$ and $\sim 0.7$ for AA and random models, respectively. We also observed that the data for the AA model with open boundary conditions and random model with periodic boundary conditions collapse better than that for the AA model with periodic boundary conditions. The critical disorder strength $\lambda_c$ for which the MBL transition happens is found to be $\approx 3t$ for the AA model with open boundary conditions, $\approx 3.4t$ for the AA model with periodic boundary conditions and $\approx 3.6t$ for the random model with periodic boundary conditions.

\section{Standard deviation of entanglement entropy}\label{sec:stdsent}
\begin{figure}[h!]
\begin{center}
\includegraphics[width=\linewidth]{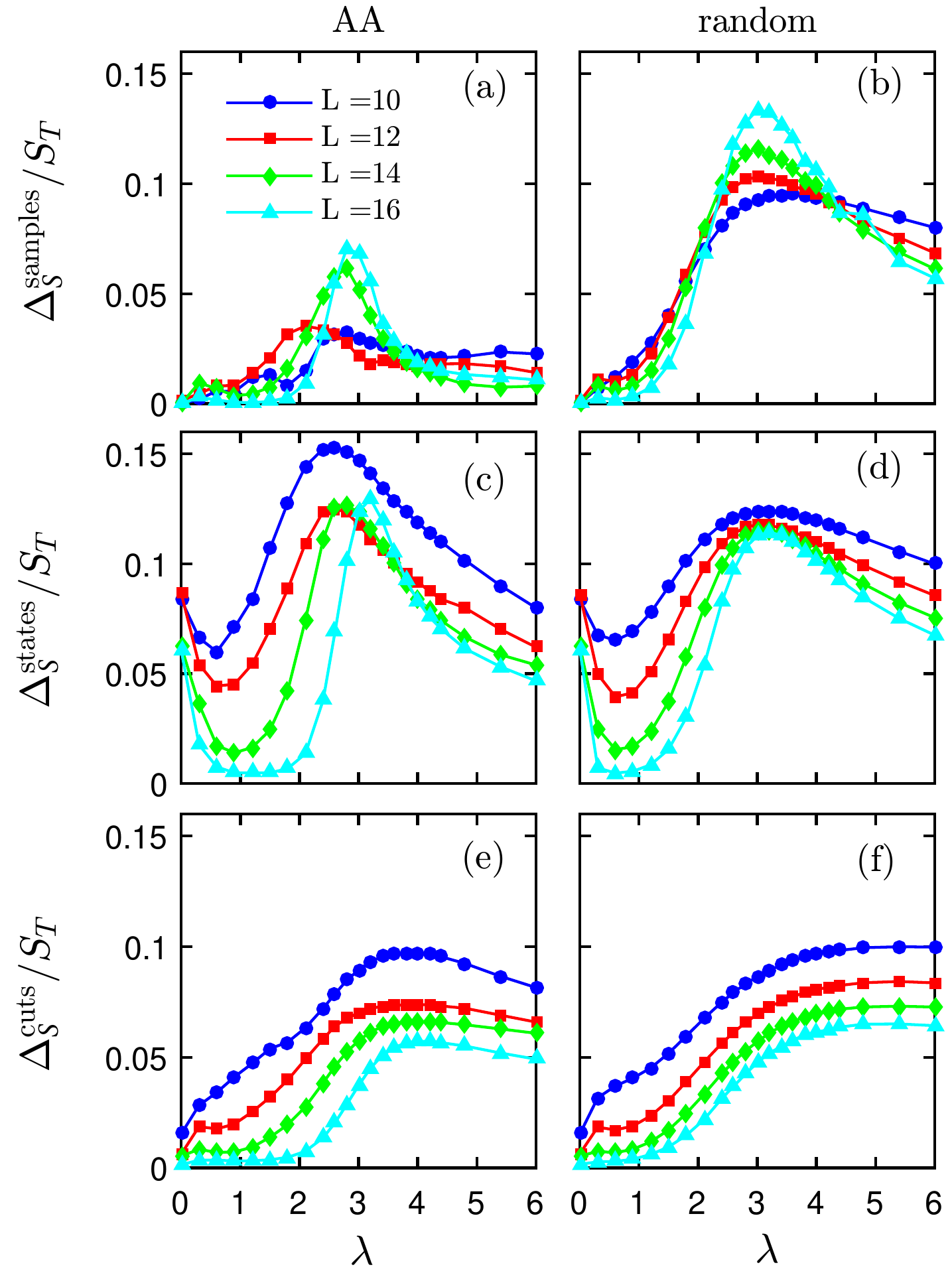}
\end{center}
\caption{Standard deviation of half-chain entanglement entropy over different samples ($\Delta_{S}^{\mathrm{samples}}$), eigenstates ($\Delta_{S}^{\mathrm{states}}$), and cuts ($\Delta_{S}^{\mathrm{cuts}}$) for AA (left panel) and random (right panel) models with different system size $L$. The standard deviation is calculated over 100 states closest to the zero energy. 
}\label{fig:fig10}
\end{figure}

In this appendix, we calculate the standard deviation of half-chain entanglement entropy over samples  $\Delta^{\mathrm{samples}}_S$, over eigenstates $\Delta^{\mathrm{states}}_S$, and over cuts  $\Delta^{\mathrm{cuts}}_S$, respectively, as shown in Fig.~\ref{fig:fig10}. Following Refs.~\cite{Khemani-2016,Khemani-2017}, we parse our data and define the three quantities above as follows: (a)  $\Delta^{\mathrm{samples}}_S$ (which denotes the sample-to-sample variation of the entanglement entropy) is calculated by first averaging the half-chain entanglement entropy over eigenstates and spatial cuts in a given sample and then computing the standard deviation of the averaged entanglement entropy over different samples, (b) $\Delta^{\mathrm{states}}_S$ (which denotes the eigenstate-to-eigenstate variation of the entanglement entropy) is computed by first averaging the half-chain entanglement entropy over all spatial cuts, then taking the standard deviation over eigenstates for a given sample and finally averaging across samples, and (c) $\Delta^{\mathrm{cut}}_S$ (which denotes the cut-to-cut variation of the entanglement entropy) is obtained by first calculating the standard deviation over spatial cuts in a given eigenstate and then averaging over the samples and eigenstates. To compute the above quantities, in this paper, we calculate the half-chain entanglement entropy by considering only 100 eigenstates closest to zero energy over different spatial cuts, eigenstates, and samples. We find results consistent with those obtained in Ref.~\cite{Khemani-2017}: $\Delta^{\mathrm{samples}}_S$ is growing superlinearly in $L$ near $\lambda_c$ for the random model and growing more weakly with $L$ in the quasiperiodic model with its magnitude much smaller than the random model.  $\Delta^{\mathrm{states}}_S/S_T$ is peaked and essentially $L$ independent near the transition for the random model while the peak is growing weakly with $L$ in the quasiperiodic case. $\Delta^{\mathrm{cuts}}_S$ obeys a subvolume law scaling in each regime of the phase diagram.

\bibliography{references}
\end{document}